\documentclass[12pt]{article}
\usepackage{amsmath,amssymb}
\usepackage{bm}
\usepackage{mathrsfs}
\usepackage{fourier}
\usepackage{multirow}
\allowdisplaybreaks[4]
\newcommand{{\SlashD}}{D\!\!\!\!\!\!\big/}
\newcommand{{\Slashq}}{q\!\!\!\!\!\big/}
\usepackage[usenames]{color}

\begin{document}

\title{Yukawa interactions, flavor symmetry, 
and non-canonical K\"{a}hler potential}

\author{%       %Use \scshape  for the family name
Yoshiharu \textsc{Kawamura}\footnote{E-mail: haru@azusa.shinshu-u.ac.jp}\\
{\it Department of Physics, Shinshu University, }\\
{\it Matsumoto 390-8621, Japan}\\
}

\date{%      %Editorial Office will fill in this.
August 20, 2018}

\maketitle
\begin{abstract}
We study the origin of fermion mass hierarchy
and flavor mixing in the standard model,
paying attention to flavor symmetries and fermion kinetic terms.
There is a possibility that the hierarchical flavor structure
of quarks and charged leptons
originates from non-canonical types of fermion kinetic terms
in the presence of flavor symmetric Yukawa interactions.
A flavor symmetry can be hidden in the form of 
non-unitary bases in the standard model.
The structure of K\"{a}hler potential can become
a touchstone of new physics.
\end{abstract}

%\newpage

\section{Introduction}

The origin of fermion mass hierarchy and flavor mixing
has been a big mystery, which comes from the fact that
there is no powerful principle to determine Yukawa couplings
in the standard model (SM).
Yukawa couplings are expressed as general square matrices 
taking complex values,
and they are diagonalized by bi-unitary transformations.
Their eigenvalues become quark and charged lepton masses
after multiplying the vacuum expectation value (VEV) of neutral component
in the Higgs doublet.
The mixing of flavors occurs from the difference
between mass eigenstates and weak interaction ones~\cite{MNS,C,KM}.

There have been many intriguing attempts to explain the values
of physical parameters concerning fermion masses and
flavor mixing matrices.
Most of them are based 
on the top-down approach~\cite{CEG,F,HHW,F2,GJ,FN}, i.e.,
Yukawa couplings are constructed or given in the form of Ansatz
based on high-energy physics 
such as grand unified theories (GUTs) and superstring theories (SSTs)
or extensions of SM with some flavor symmetry,
and the analyses have been carried out
model-dependently and/or independently 
from the phenomenological point of view.

At present, any evidences from new physics 
except for neutrinos
have not yet been discovered,
and new physics might be beyond all imagination.
Hence, it would be interesting to see flavor physics
through a different lens, with the expectation 
that it offers some hints of a fundamental theory.
We adopt several reasonable assumptions in a theory beyond the SM.
(a) {\it The field variables are not necessarily the same
as those in the SM.}
(b) {\it There is a symmetry relating to the flavor or family of the SM
(a flavor or family symmetry).}
{\it The symmetry is broken down by the VEVs of some scalar fields called flavons.}
(c) {\it Flavons couple to matter fields through matter kinetic terms
dominantly.}
The second assumption is based on the idea that
the family number is naturally understood
as a dimension of representation and
a predictability is improved by the reduction of free parameters,
in the presence of a flavor or family symmetry.
The last one is based on the fact 
that various fields are easy to couple to among them
in a K\"{a}hler potential,
compared with a superpotential
controlled by holomorphy,
and the K\"{a}hler potential can change by receiving radiative corrections
in contrast with the superpotential,
in supersymmetric (SUSY) theories.
We expect that the SUSY exists
in an underlying theory,
even if it is broken down at a high energy scale.

Suppose that a flavor symmetry exist,
we have several questions such as
``what type of symmetry exists?'',
``what is the breaking mechanism ?''
and ``how is it hidden in the SM?''.
Here, we focus interest on the last one.
There is a possibility that a flavor symmetry is hidden
in the form of non-unitary bases, i.e.,
matter fields in the SM are transformed by non-unitary matrices.
In Appendix A, we give an illustration
of a realization of U$(N)$ symmetry using non-unitary matrices.

Our approach is summarized as follows.
We suppose field variables respecting a flavor symmetry
(that the corresponding transformation is realized by unitary matrices)
and rewrite the Lagrangian density in the SM 
using such variables.
We investigate the structure of terms violating 
the flavor symmetry,
and attempt to conjecture physics beyond the SM.
Although physics is unchanged by a choice of field variables
and representations,
there can be a difference in an understandability of physical phenomena.
For instance, in the relativistic quantum mechanics,
the Dirac representation of $\gamma$ matrices 
is useful to analyze non-relativistic phenomena
and the chiral representation is suitable to investigate high-energy physics.
It is desirable to find helpful field variables in order to envisage
a mechanism of flavor symmetry breaking in an underlying theory.
We expect that unitary bases of flavor symmetries
are suitable to describe physics right after
the breakdown of flavor symmetries,
although they are unfit for perturbative calculations due to 
the presence of non-canonical kinetic terms.
One of the best plans would be to attack a flavor structure
from both bottom-up and top-down approaches.
Knowledge and information obtained by the bottom-up approach
can provide a new procedure based on a top-down approach.

In this paper, we study the origin of fermion mass hierarchy
and flavor mixing in the SM, using the above-mentioned approach.
We examine whether the hierarchical flavor structure
of quarks and charged leptons
can originate from specific forms of their kinetic terms
in the presence of flavor symmetric Yukawa interactions or not.
We also propose a variant procedure based on the top-down approach.

The outline of this paper is as follows.
In the next section, we review quark Yukawa interactions 
and a no-go theorem on flavor symmetries
in the SM.
We explore the origin of the hierarchical structure of quarks 
and charged leptons, paying attention to flavor symmetries 
and fermion kinetic terms in Sect. 3.
In the last section, we give conclusions and discussions.

\section{Yukawa interactions and flavor symmetry}

We review quark Yukawa interactions 
and the absence of exact flavor symmetries in the SM.

\subsection{Quark Yukawa interactions}

Let us start with the Lagrangian densities of the quark sector,
\begin{eqnarray}
&~& \mathscr{L}_{\rm kinetic}^{\rm quark} = \overline{q}_{{\rm L}i} i \SlashD q_{{\rm L}i}
+ \overline{u}_{{\rm R}i} i \SlashD u_{{\rm R}i}
+ \overline{d}_{{\rm R}i} i \SlashD d_{{\rm R}i},
\label{L-kin-quark}\\
&~& \mathscr{L}_{\rm Yukawa}^{\rm quark} 
= - y_{ij}^{(u)} \overline{q}_{{\rm L}i} \tilde{\phi} u_{{\rm R}j}
- y_{ij}^{(d)} \overline{q}_{{\rm L}i} \phi d_{{\rm R}j} + {\rm h.c.},
\label{L-Y-quark}
\end{eqnarray}
where $q_{{\rm L}i}$ are left-handed quark doublets,
$u_{{\rm R}i}$ and $d_{{\rm R}i}$ are
right-handed up- and down-type quark singlets,
$i, j (=1, 2, 3)$ are family labels, 
summation over repeated indices is understood throughout  this paper,
$y_{ij}^{(u)}$ and $y_{ij}^{(d)}$ are Yukawa couplings,
$\phi$ is the Higgs doublet, $\tilde{\phi} = i \tau_2 \phi^*$
and h.c. stands for hermitian conjugation of former terms.
The Yukawa couplings are diagonalized 
as $V_{\rm L}^{(u)} y^{(u)} {V_{\rm R}^{(u)}}^{\dagger} = y_{\rm diag}^{(u)}$
and $V_{\rm L}^{(d)} y^{(d)} {V_{\rm R}^{(d)}}^{\dagger} = y_{\rm diag}^{(d)}$
by bi-unitary transformations 
and the quark masses are obtained as
\begin{eqnarray}
&~& V_{\rm L}^{(u)} y^{(u)} {V_{\rm R}^{(u)}}^{\dagger} \frac{v}{\sqrt{2}}
= y_{\rm diag}^{(u)} \frac{v}{\sqrt{2}} = M_{\rm diag}^{(u)} = {\rm diag}\left(m_u, m_c, m_t\right),
\label{Mu-diag}\\
&~& V_{\rm L}^{(d)} y^{(d)} {V_{\rm R}^{(d)}}^{\dagger} \frac{v}{\sqrt{2}}
= y_{\rm diag}^{(d)} \frac{v}{\sqrt{2}} = M_{\rm diag}^{(d)} = {\rm diag}\left(m_d, m_s, m_b\right),
\label{Md-diag}
\end{eqnarray}
where $V_{\rm L}^{(u)}$, $V_{\rm L}^{(d)}$, $V_{\rm R}^{(u)}$ and $V_{\rm R}^{(d)}$
are unitary matrices, $v/\sqrt{2}$ is the VEV of neutral component 
in the Higgs doublet, family labels are omitted,
and $m_u$, $m_c$, $m_t$, $m_d$, $m_s$ and $m_b$ are masses of up, charm, top,
down, strange and bottom quarks, respectively.

The Yukawa couplings are expressed by
\begin{eqnarray}
y^{(u)} = {V_{\rm L}^{(u)}}^{\dagger} y_{\rm diag}^{(u)} V_{\rm R}^{(u)},~~
y^{(d)} = {V_{\rm L}^{(d)}}^{\dagger} y_{\rm diag}^{(d)} V_{\rm R}^{(d)}
= {V_{\rm L}^{(u)}}^{\dagger} V_{\rm KM} y_{\rm diag}^{(d)} V_{\rm R}^{(d)},
\label{yd}
\end{eqnarray}
using $V_{\rm L}^{(u)}$, $V_{\rm R}^{(u)}$, $V_{\rm R}^{(d)}$,
$y_{\rm diag}^{(u)}$, $y_{\rm diag}^{(d)}$,
and the Kobayashi-Maskawa matrix defined by~\cite{KM}
\begin{eqnarray}
V_{\rm KM} \equiv V_{\rm L}^{(u)} {V_{\rm L}^{(d)}}^{\dagger}.
\label{VKM}
\end{eqnarray}
Information on physics beyond the SM is hidden
in $V_{\rm L}^{(u)}$,
$V_{\rm R}^{(u)}$, and $V_{\rm R}^{(d)}$ besides 
observable parameters $y_{\rm diag}^{(u)}$, $y_{\rm diag}^{(d)}$,
and $V_{\rm KM}$.
The matrices $V_{\rm L}^{(u)}$,
$V_{\rm R}^{(u)}$, and $V_{\rm R}^{(d)}$
are completely unknown in the SM,
because they can be eliminated by the global 
${\rm U}(3) \times {\rm U}(3) \times {\rm U}(3)/{\rm U}(1)$ symmetry
that the quark kinetic term $\mathscr{L}_{\rm kinetic}^{\rm quark}$ possesses.

From (\ref{Mu-diag}), (\ref{Md-diag}) and experimental values of quark masses,
the eigenvalues of $y^{(u)}$ and $y^{(d)}$ are roughly estimated 
at the weak scale as
\begin{eqnarray}
&~& y_{\rm diag}^{(u)} \doteqdot {\rm diag}\left(1.3 \times 10^{-5},~ 7.3 \times 10^{-3},~ 1.0\right),
\label{yu-diag-value}\\
&~& y_{\rm diag}^{(d)} \doteqdot {\rm diag}\left(2.7 \times 10^{-5},~ 
5.5 \times 10^{-4},~ 2.4 \times 10^{-2}\right).
\label{yd-diag-value}
\end{eqnarray}
We find that there is a large hierarchy among a size of Yukawa couplings, 
and it has thrown up the big mystery of its origin.
From (\ref{yd}), we derive the relation:
\begin{eqnarray}
y^{(d)} {V_{\rm R}^{(d)}}^{\dagger}
= y^{(u)} {V_{\rm R}^{(u)}}^{\dagger}
y_{\rm diag}^{(u)-1} V_{\rm KM} y_{\rm diag}^{(d)},
\label{yV}
\end{eqnarray}
where $y_{\rm diag}^{(u)-1}$is the inverse matrix of $y_{\rm diag}^{(u)}$.
The matrix $y_{\rm diag}^{(u)-1} V_{\rm KM} y_{\rm diag}^{(d)}$
can be a barometer of the difference 
between $y^{(u)} {V_{\rm R}^{(u)}}^{\dagger}$ and $y^{(d)} {V_{\rm R}^{(d)}}^{\dagger}$,
and it is roughly estimated at the weak scale as
\begin{eqnarray}
&~& y_{\rm diag}^{(u)-1} V_{\rm KM} y_{\rm diag}^{(d)}
\doteqdot \left(
\begin{array}{ccc}
\left(1-\frac{\lambda^2}{2}\right) \frac{m_d}{m_u} & \lambda \frac{m_s}{m_u} 
& A \lambda^3 (\rho - i \eta) \frac{m_b}{m_u}\\
- \lambda \frac{m_d}{m_c} & \left(1-\frac{\lambda^2}{2}\right) \frac{m_s}{m_c}
& A \lambda^2 \frac{m_b}{m_c}\\
A \lambda^3 (1 - \rho - i \eta) \frac{m_d}{m_t} & - A \lambda^2 \frac{m_s}{m_t}
& \frac{m_b}{m_t}
\end{array}
\right)
\nonumber \\
&~& ~~~~~~~~~~~~~~~~~~~~~~~~~~~~~~
= \left(
\begin{array}{ccc}
O(1) & O(10) & O(10)\\
O\left(10^{-3}\right) & O\left(10^{-1}\right) & O\left(10^{-1}\right)\\
O\left(10^{-7}\right) & O\left(10^{-5}\right) & O\left(10^{-2}\right)
\end{array}
\right),
\label{yVy}
\end{eqnarray}
where  we use the Wolfenstein parametrization~\cite{Wolf}, i.e.,
$\lambda = \sin\theta_{\rm C} \doteqdot 0.225$ 
($\theta_{\rm C}$ is the Cabibbo angle~\cite{C}),
$A \doteqdot 0.811$, $\rho$ and $\eta$ are real parameters~\cite{PDG}.

\subsection{No unbroken flavor symmetry}

We explain that there is no unbroken flavor-dependent symmetry
respecting the ${\rm SU}(2)_{\rm L}$ gauge symmetry~\cite{LNS,Koide}.
If the quark sector is invariant under a global transformation
(a low-energy remnant of some flavor symmetries):
\begin{eqnarray}
q_{\rm L} \to F_{\rm L} q_{\rm L},~~
u_{\rm R} \to F_{\rm R}^{(u)} u_{\rm R},~~d_{\rm R} \to F_{\rm R}^{(d)} d_{\rm R},~~
\phi \to e^{i \theta} \phi,
\label{F-tr}
\end{eqnarray}
the quark Yukawa couplings should satisfy the relations:
\begin{eqnarray}
e^{-i\theta} F_{\rm L}^{\dagger} y^{(u)} F_{\rm R}^{(u)} = y^{(u)},~~
e^{i\theta} F_{\rm L}^{\dagger} y^{(d)} F_{\rm R}^{(d)} = y^{(d)},
\label{F-inv}
\end{eqnarray}
where $F_{\rm L}$, $F_{\rm R}^{(u)}$, and $F_{\rm R}^{(d)}$
are $3 \times 3$ unitary matrices, and $\theta$ is a real number.
From (\ref{F-inv}), we have the relations:
\begin{eqnarray}
&~& \left[y^{(u)} {y^{(u)}}^{\dagger},~ F_{\rm L}\right] = 0,~~
\left[{y^{(u)}}^{\dagger} y^{(u)},~ F_{\rm R}^{(u)}\right] = 0,
\label{yu-com}\\
&~& \left[y^{(d)} {y^{(d)}}^{\dagger},~ F_{\rm L}\right] = 0,~~
\left[{y^{(d)}}^{\dagger} y^{(d)},~ F_{\rm R}^{(d)}\right] = 0,
\label{yd-com}
\end{eqnarray}
and then $F_{\rm L}$ can also be diagonalized by the unitary matrices
$V_{\rm L}^{(u)}$ and $V_{\rm L}^{(d)}$
which diagonalize $y^{(u)} {y^{(u)}}^{\dagger}$ and $y^{(d)} {y^{(d)}}^{\dagger}$
such that
\begin{eqnarray}
V_{\rm L}^{(u)} F_{\rm L} {V_{\rm L}^{(u)}}^{\dagger} = F_{\rm L~diag}^{(u)},~~
V_{\rm L}^{(d)} F_{\rm L} {V_{\rm L}^{(d)}}^{\dagger} = F_{\rm L~diag}^{(d)}.
\label{FLdiag}
\end{eqnarray}
In the same way, $F_{\rm R}^{(u)}$ and $F_{\rm R}^{(d)}$ can also be
diagonalized by the unitary matrices $V_{\rm R}^{(u)}$ and $V_{\rm R}^{(d)}$
which diagonalize ${y^{(u)}}^{\dagger} y^{(u)}$ and ${y^{(d)}}^{\dagger} y^{(d)}$
such that
\begin{eqnarray}
V_{\rm R}^{(u)} F_{\rm R}^{(u)} {V_{\rm R}^{(u)}}^{\dagger} = F_{\rm R~diag}^{(u)},~~
V_{\rm R}^{(d)} F_{\rm R}^{(d)} {V_{\rm R}^{(d)}}^{\dagger} = F_{\rm R~diag}^{(d)}.
\label{FRdiag}
\end{eqnarray}

By multiplying both sides of each relation in (\ref{F-inv})
by $V_{\rm L}^{(u)}$ and $V_{\rm L}^{(d)}$ from the left 
and ${V_{\rm R}^{(u)}}^{\dagger}$ and ${V_{\rm R}^{(d)}}^{\dagger}$ from the right
and using (\ref{yd}), (\ref{FLdiag}) and (\ref{FRdiag}),
the following relations are obtained,
\begin{eqnarray}
e^{-i\theta} F_{\rm L~diag}^{(u) \dagger} y_{\rm diag}^{(u)} F_{\rm R~diag}^{(u)} 
= y_{\rm diag}^{(u)},~~
e^{i\theta} F_{\rm L~diag}^{(d) \dagger} y_{\rm diag}^{(d)} F_{\rm R~diag}^{(d)} 
= y_{\rm diag}^{(d)},
\label{F-inv-diag}
\end{eqnarray}
and they lead to $F_{\rm L~diag}^{(u)} = e^{-i\theta} F_{\rm R~diag}^{(u)}$
and $F_{\rm L~diag}^{(d)} = e^{i\theta} F_{\rm R~diag}^{(d)}$.

From (\ref{FLdiag}), we obtain the relation:
\begin{eqnarray}
F_{\rm L~diag}^{(u)} V_{\rm KM} = V_{\rm KM} F_{\rm L~diag}^{(d)}.
\label{VF=FV}
\end{eqnarray}
Then, we find that $F_{\rm L~diag}^{(u)} = F_{\rm L~diag}^{(d)} = e^{i\varphi} I$
(where $\varphi$ is a real number and $I$ is the $3 \times 3$ identity matrix)
from the fact that all mixing angles of $V_{\rm KM}$ are nonzero,
and it means that any exact flavor-dependent symmetries do not exist
in the quark sector of the SM.
In the same way, it is shown that any exact flavor-dependent symmetries
do not also survive in the lepton sector of the SM.

\section{K\"{a}hler structure in SM and beyond}

Based on feasible assumptions in a theory beyond the SM
such that the field variables are not necessarily the same
as those in the SM,
there is a flavor symmetry broken down by the VEVs of flavons
and flavons couple to matter fields in matter kinetic terms
dominantly, we rewrite the Lagrangian density in the SM 
using unitary bases of a flavor symmetry,
investigate the structure of terms violating the flavor symmetry,
and attempt to conjecture physics beyond the SM.
Here, unitary bases mean sets of fields that are transformed by unitary matrices.
For more details, see Appendix A.

\subsection{Change of variables and matching conditions}

We assume that a theory beyond the SM has
a flavor symmetry\footnote{
The flavor structure of quarks and leptons has been studied
intensively, based on various flavor symmetries~\cite{FN,MY,I,HPS,HS,AF,IKOSOT,IKOOST}.
}
and the symmetry is broken down by the VEVs of flavons
at some high-energy scale near $M_{\rm BSM}$.
Here, $M_{\rm BSM}$ is an energy scale of new physics or
the upper limit of a scale where the SM holds.
We assume that $M_{\rm BSM}$ is much bigger than the weak scale, for simplicity.
In this case, there is a possibility that
we obtain useful information on flavor physics
from the matching conditions at $M_{\rm BSM}$.

We denote unitary bases of a flavor group ${\rm G}_{\rm F}$ for quarks
by $q'_{\rm L}$, $u'_{\rm R}$, and $d'_{\rm R}$.
They transform as
\begin{eqnarray}
q'_{\rm L} \to F_{\rm L} q'_{\rm L},~~
u'_{\rm R} \to F_{\rm R}^{(u)} u'_{\rm R},~~d'_{\rm R} \to F_{\rm R}^{(d)} d'_{\rm R},~~
\phi \to e^{i \theta} \phi,
\label{F-tr-prime}
\end{eqnarray}
under the ${\rm G}_{\rm F}$ transformation,
where $F_{\rm L}$, $F_{\rm R}^{(u)}$, and $F_{\rm R}^{(d)}$ are $3 \times 3$ unitary matrices.
Then, the Yukawa interaction terms are rewritten as
\begin{eqnarray}
\mathscr{L}_{\rm Yukawa}^{\rm quark} 
= - \left(y_1\right)_{ij} \overline{q}'_{{\rm L}i} \tilde{\phi} u'_{{\rm R}j}
- \left(y_2\right)_{ij} \overline{q}'_{{\rm L}i} \phi d'_{{\rm R}j} + {\rm h.c.},
\label{L-Y-quark-prime}
\end{eqnarray}
where $\left(y_1\right)_{ij}$ and $\left(y_2\right)_{ij}$ are 
Yukawa couplings in the unitary bases of flavor symmetry.
These couplings, in general, consist of two parts, i.e.,
$\left(y_1\right)_{ij} = \left(y_1^{\rm F}\right)_{ij} + \left(\varDelta y_1\right)_{ij}$
and $\left(y_2\right)_{ij} = \left(y_2^{\rm F}\right)_{ij} + \left(\varDelta y_2\right)_{ij}$.
Here, $\left(y_1^{\rm F}\right)_{ij}$ and $\left(y_2^{\rm F}\right)_{ij}$ are 
${\rm G}_{\rm F}$-invariant couplings satisfying
$e^{-i\theta} F_{\rm L}^{\dagger} y_1^{\rm F} F_{\rm R}^{(u)} = y_1^{\rm F}$
and $e^{i\theta} F_{\rm L}^{\dagger} y_2^{\rm F} F_{\rm R}^{(d)} = y_2^{\rm F}$, respectively,
and $\left(\varDelta y_1\right)_{ij}$ and $\left(\varDelta y_2\right)_{ij}$ are non-invariant ones
showing the breakdown of ${\rm G}_{\rm F}$ due to the VEVs of flavons.

The unitary bases of ${\rm G}_{\rm F}$ are related to 
the SM ones $q_{\rm L}$, $u_{\rm R}$, and $d_{\rm R}$ by the change of variables as
\begin{eqnarray}
&~& q_{\rm L} = V_q J_q  U_q q'_{\rm L},~~ \left(\overline{q}_{\rm L} 
= \overline{q}'_{\rm L}U_q^{\dagger}J_q V_q^{\dagger}\right),
\label{q-q'}\\
&~& u_{\rm R} = \left(y^{(u)}\right)^{-1} V_q J_q^{-1} U_q y_1 u'_{\rm R}
= {V_{\rm R}^{(u)}}^{\dagger} \left(y_{\rm diag}^{(u)}\right)^{-1} V_{\rm L}^{(u)}
V_q J_q^{-1} U_q y_1 u'_{\rm R},~~
\label{u-u'}\\
&~& d_{\rm R} = \left(y^{(d)}\right)^{-1} V_q J_q^{-1} U_q y_2 d'_{\rm R}
= {V_{\rm R}^{(d)}}^{\dagger} \left(y_{\rm diag}^{(d)}\right)^{-1} V_{\rm KM}^{\dagger} V_{\rm L}^{(u)}
V_q J_q^{-1} U_q y_2 d'_{\rm R},~~
\label{d-d'}
\end{eqnarray}
where $V_q$ and $U_q$ are $3 \times 3$ unitary matrices and
$J_q$ is a real $3 \times 3$ diagonal matrix.

Using new variables, the quark kinetic terms in the SM are rewritten as
\begin{eqnarray}
\mathscr{L}_{\rm kinetic}^{\rm quark} = k_{ij}^{(q)} \overline{q}'_{{\rm L}i} i \SlashD q'_{{\rm L}j}
+ k_{ij}^{(u)} \overline{u}'_{{\rm R}i} i \SlashD u'_{{\rm R}j}
+ k_{ij}^{(d)} \overline{d}'_{{\rm R}i} i \SlashD d'_{{\rm R}j},
\label{L-kin-quark-prime}
\end{eqnarray}
where the kinetic coefficients $k_{ij}^{(q)}$, $k_{ij}^{(u)}$, and $k_{ij}^{(d)}$ are given by
\begin{eqnarray}
&~& k_{ij}^{(q)} = \left(U_q^{\dagger} (J_{q})^2 U_q\right)_{ij},
\label{kq-GF}\\
&~& k_{ij}^{(u)} = \left(y_1^{\dagger}{W^{(u)}}^{\dagger} 
\left(y_{\rm diag}^{(u) -1}\right)^2 W^{(u)} y_1\right)_{ij},
\label{ku-GF}\\
&~& k_{ij}^{(d)} = \left(y_2^{\dagger}{W^{(d)}}^{\dagger} 
\left(y_{\rm diag}^{(d) -1}\right)^2 W^{(d)} y_2\right)_{ij}
= \left(y_2^{\dagger}{W^{(u)}}^{\dagger} 
V_{\rm KM} \left(y_{\rm diag}^{(d) -1}\right)^2 
V_{\rm KM}^{\dagger} W^{(u)} y_2\right)_{ij}.
\label{kd-GF}
\end{eqnarray}
Here, $W^{(u)} = V_{\rm L}^{(u)} V_q J_{q}^{-1} U_q$, 
$W^{(d)} = V_{\rm L}^{(d)} V_q J_{q}^{-1} U_q$,
and we use the feature that the kinetic coefficients are positive definite.
Note that $W^{(u)} $ and $W^{(d)}$ are not necessarily unitary matrices.
If $J_q$ is the identity matrix,
$k_{ij}^{(q)}$ is the canonical one (the identity matrix)
and $W^{(u)} $ and $W^{(d)}$ become unitary matrices.

We give an alternative proof on the absence of exact flavor symmetries in the SM briefly.
Under the assumption that $\mathscr{L}_{\rm Yukawa}^{\rm quark}$ 
given in (\ref{L-Y-quark-prime}) is invariant under the transformation (\ref{F-tr-prime}), i.e.,
$e^{-i\theta} F_{\rm L}^{\dagger} y_1 F_{\rm R}^{(u)} = y_1$
and $e^{i\theta} F_{\rm L}^{\dagger} y_2 F_{\rm R}^{(d)} = y_2$,
it is shown that no exact flavor symmetries exist from the invariance of
$\mathscr{L}_{\rm kinetic}^{\rm quark}$ 
given in (\ref{L-kin-quark-prime}) under the transformation (\ref{F-tr-prime}),
in the following.
Eigenvalues of $F_{\rm R}^{(u)}$ and $F_{\rm R}^{(d)}$
are given by those of $F_{\rm L}$ multiplied by $e^{i\theta}$
and $e^{-i\theta}$, respectively, as estimated from (\ref{F-inv-diag}).
Using  (\ref{kq-GF}) and $F_{\rm L}^{\dagger} k^{(q)} F_L = k^{(q)}$
derived from the invariance of $k_{ij}^{(q)} \overline{q}'_{{\rm L}i} i \SlashD q'_{{\rm L}j}$,
we find that $F_{\rm L}$ is diagonalized by $U_q$ 
as $U_q F_{\rm L} U_q^{\dagger} = F_{\rm L~diag}$.
Here, we omit the labels of flavor.
From (\ref{ku-GF}), (\ref{kd-GF}), and ${F_{\rm R}^{(u)}}^{\dagger} k^{(u)} F_{\rm R}^{(u)} = k^{(u)}$
and ${F_{\rm R}^{(d)}}^{\dagger} k^{(d)} F_{\rm R}^{(d)} = k^{(d)}$
derived from the invariance of other kinetic terms,
we obtain the relations $\tilde{F}_{\rm L~diag} V_{\rm L}^{(u)} = V_{\rm L}^{(u)} F_{\rm L~diag}$
and $\tilde{F}_{\rm L~diag} V_{\rm L}^{(d)} = V_{\rm L}^{(d)} F_{\rm L~diag}$,
using $e^{-i\theta} F_{\rm L}^{\dagger} y_1 F_{\rm R}^{(u)} = y_1$,
$e^{i\theta} F_{\rm L}^{\dagger} y_2 F_{\rm R}^{(d)} = y_2$, and
$U_q F_{\rm L} U_q^{\dagger} = F_{\rm L~diag}$.
Here, $\tilde{F}_{\rm L~diag}$ is a diagonal unitary matrix.
These relations lead to $\tilde{F}_{\rm L~diag} V_{\rm KM} = V_{\rm KM} \tilde{F}_{\rm L~diag}$
which means that $\tilde{F}_{\rm L~diag}$ and ${F}_{\rm L~diag}$
should be proportional to the identity matrix
or the non-existence of exact flavor-dependent symmetries.

From (\ref{yu-diag-value}) and (\ref{yd-diag-value}),
$\left(y_{\rm diag}^{(u) -1}\right)^2$, $\left(y_{\rm diag}^{(d) -1}\right)^2$
and $V_{\rm KM} \left(y_{\rm diag}^{(d) -1}\right)^2 
V_{\rm KM}^{\dagger}$ are roughly estimated at the weak scale as
\begin{eqnarray}
&~& \left(y_{\rm diag}^{(u) -1}\right)^2 
\doteqdot {\rm diag}\left(5.9 \times 10^{9},~ 1.9 \times 10^{4},~ 1.0\right)
\nonumber \\
&~& ~~~~~~~~~~~~~~~~~~ 
= 5.9 \times 10^{9} \times {\rm diag}\left(1,~ 3.2 \times 10^{-6},~ 1.7 \times 10^{-10}\right),
\label{yu^(-2)-diag-value}\\
&~& \left(y_{\rm diag}^{(d) -1}\right)^2 
\doteqdot {\rm diag}\left(1.4 \times 10^{9},~ 3.8 \times 10^{6},~ 2.1 \times 10^{3}\right)
\nonumber \\
&~& ~~~~~~~~~~~~~~~~~~ 
= 1.4 \times 10^{9} \times {\rm diag}\left(1,~ 2.7 \times 10^{-3},~ 1.5 \times 10^{-6}\right),
\label{yd^(-2)-diag-value}\\
&~& V_{\rm KM} \left(y_{\rm diag}^{(d) -1}\right)^2 V_{\rm KM}^{\dagger}
\doteqdot 1.4 \times 10^{9} \times \left(
\begin{array}{ccc}
1-\lambda^2 & -\lambda & O\left(\lambda^3\right) \\
-\lambda & \lambda^2 & O\left(\lambda^4\right) \\
O\left(\lambda^3\right) & O\left(\lambda^4\right) & O\left(\lambda^6\right)
\end{array} 
\right).
\label{VyV}
\end{eqnarray}
Physical parameters, in general, receive radiative corrections,
and the above values should be evaluated by considering 
renormalization effects 
and should match with their counterparts at $M_{\rm BSM}$.
From (\ref{L-Y-quark-prime}) and (\ref{L-kin-quark-prime}),
information on the flavor structure in the SM
is transfered to $k_{ij}^{(q)}$, $k_{ij}^{(u)}$ and $k_{ij}^{(d)}$
in the kinetic terms.

To speculate a theory of quarks beyond the SM,
let us describe it by
\begin{eqnarray}
&~& \mathscr{L}_{\rm BSM}^{\rm quark} 
= K_{ij}^{(q)} \overline{q}'_{{\rm L}i} i \SlashD q'_{{\rm L}j}
+ K_{ij}^{(u)} \overline{u}'_{{\rm R}i} i \SlashD u'_{{\rm R}j}
+ K_{ij}^{(d)} \overline{d}'_{{\rm R}i} i \SlashD d'_{{\rm R}j}
\nonumber \\
&~&  ~~~~~~~~~~~~~~~~~~ - \left(Y_1\right)_{ij} \overline{q}'_{{\rm L}i} \tilde{\phi} u'_{{\rm R}j}
- \left(Y_2\right)_{ij} \overline{q}'_{{\rm L}i} \phi d'_{{\rm R}j} + {\rm h.c.},
\label{L-BSM-quark}
\end{eqnarray}
where $K_{ij}^{(q)}$, $K_{ij}^{(u)}$, $K_{ij}^{(d)}$, 
$\left(Y_1\right)_{ij}$, and $\left(Y_2\right)_{ij}$
contain fields such that
$\mathscr{L}_{\rm BSM}^{\rm quark}$ is invariant under
the G$_{\rm F}$ transformation.
The $\mathscr{L}_{\rm BSM}^{\rm quark}$
describes only a part relating to quarks in new physics,
and chiral anomalies are supposed to be canceled by other contributions
if the G$_{\rm F}$ symmetry is local.

When $\mathscr{L}_{\rm kinetic}^{\rm quark}$
and $\mathscr{L}_{\rm Yukawa}^{\rm quark}$
are obtained by getting the VEVs
after the breakdown of G$_{\rm F}$ symmetry,
the following matching conditions should be imposed on
\begin{eqnarray}
k_{ij}^{(q)} = \left\langle K_{ij}^{(q)} \right\rangle, ~~
k_{ij}^{(u)} = \left\langle K_{ij}^{(u)} \right\rangle,~~ 
k_{ij}^{(d)} = \left\langle K_{ij}^{(d)} \right\rangle,~~
\left(y_1\right)_{ij} = \left\langle \left(Y_1\right)_{ij} \right\rangle,~~
\left(y_2\right)_{ij} = \left\langle \left(Y_2\right)_{ij} \right\rangle,
\label{<KY>}
\end{eqnarray}
at $M_{\rm BSM}$, from  (\ref{L-Y-quark-prime}), (\ref{L-kin-quark-prime})
and (\ref{L-BSM-quark}).

\subsection{Examples}

As we have few hints on a flavor symmetry,
we study two examples, i.e., a case with a U(3) symmetry
and that with an S$_3$ one.
Here ${\rm S}_3$ is the permutation group
of order 3

\subsubsection{U(3) case}

In case that a U(3) family symmetry is hidden in the SM,
the Yukawa interactions are written by
$\mathscr{L}_{\rm Yukawa}^{\rm quark} = - y_1 \overline{q}'_{{\rm L}i} \tilde{\phi} u'_{{\rm R}i}
- y_2 \overline{q}'_{{\rm L}i} \phi d'_{{\rm R}i} + {\rm h.c.}$,
where $y_1$ and $y_2$ are complex numbers.
We assume that U(3) symmetric terms dominate
in Yukawa interactions.
It is justified, in case that $M_{\rm BSM}$
is much bigger than the weak scale, other terms including fermions contain
non-renormalizable higher-dimensional operators
and they can be suppressed by a power of $M_{\rm BSM}$.

Now, we conjecture a structure of K\"{a}hler metric,
based on (\ref{kq-GF}) -- (\ref{VyV}).
There are many possibilities to realize 
the quark masses and flavor mixing consistent with experimental data.
For simplicity, we assume that $J_q = I$, i.e., $k_{ij}^{(q)} = \delta_{ij}$.
Then, $k_{ij}^{(u)}$ and $k_{ij}^{(d)}$ are written by
\begin{eqnarray}
k_{ij}^{(u)} = \left|y_1\right|^2 \left({U_{\rm L}^{(u)}}^{\dagger} 
\left(y_{\rm diag}^{(u) -1}\right)^2 U_{\rm L}^{(u)}\right)_{ij},~~
k_{ij}^{(d)} 
= \left|y_2\right|^2 \left({U_{\rm L}^{(u)}}^{\dagger} 
V_{\rm KM} \left(y_{\rm diag}^{(d) -1}\right)^2 
V_{\rm KM}^{\dagger} U_{\rm L}^{(u)}\right)_{ij},
\label{kud-tilde}
\end{eqnarray}
where $U_{\rm L}^{(u)}$ is a unitary matrix.
The $U_{\rm L}^{(u)}$ is written by $U_{\rm L}^{(u)} \equiv V_{\rm L}^{(u)} V_q$
where $V_q$ is a unitary matrix reflecting the U(3) invariance
of $q_{\rm L}$'s kinetic term.
There is a possibility that $k_{ij}^{(u)}$ and $k_{ij}^{(d)}$ take
forms whose every component has an almost same magnitude of $O(1)$,
if $\displaystyle{\left|y_1\right|^2 = O\left(10^{-10}\right)}$ 
and $\displaystyle{\left|y_2\right|^2 = O\left(10^{-9}\right)}$.
It is suggested from the formulas
\begin{eqnarray}
U
\left\{
\left(
\begin{array}{ccc}
1 & 1 & 1 \\
1 & 1 & 1 \\
1 & 1 & 1
\end{array} 
\right)
+\varepsilon_1 
\left(
\begin{array}{ccc}
1 & \overline{\omega} & \omega \\
\omega & 1 & \overline{\omega} \\
\overline{\omega} & \omega & 1 
\end{array}
\right)
+\varepsilon_2 
\left(
\begin{array}{ccc}
1 & \omega & \overline{\omega} \\
\overline{\omega} & 1 & \omega \\
\omega & \overline{\omega} & 1 
\end{array}
\right)
\right\} U^{\dagger}
= 3\left(
\begin{array}{ccc}
1 & 0 & 0 \\
0 & \varepsilon_1 & 0 \\
0 & 0 & \varepsilon_2
\end{array} 
\right),
\label{democ1}
\end{eqnarray}
and 
\begin{eqnarray}
\hspace{-1.8cm}
&~& U
\left\{
\left(
\begin{array}{ccc}
1 & 1 & 1 \\
1 & 1 & 1 \\
1 & 1 & 1
\end{array} 
\right)
+\lambda 
\left(
\begin{array}{ccc}
-2  & \omega & \overline{\omega} \\
\overline{\omega} & 1 & -2{\omega} \\
{\omega} & -2\overline{\omega} & 1 
\end{array}
\right)
+\lambda^2 
\left(
\begin{array}{ccc}
0 & \overline{\omega}-1 & \omega-1 \\
\omega-1 & 0 & \overline{\omega}-1 \\
\overline{\omega}-1 & \omega-1 & 0 
\end{array}
\right)
\right\} U^{\dagger}
\nonumber \\
\hspace{-1.8cm}&~& ~~~~~~~~~ = 3\left(
\begin{array}{ccc}
1-\lambda^2 & -\lambda & 0 \\
-\lambda & \lambda^2 & 0 \\
0 & 0 & 0
\end{array} 
\right)
\label{democ2}
\end{eqnarray}
with the unitary matrix
\begin{eqnarray}
U = \frac{1}{\sqrt{3}}
\left(
\begin{array}{ccc}
1 & 1 & 1 \\
1 & \overline{\omega} & \omega \\
1 & \omega & \overline{\omega}
\end{array}
\right),
\label{U}
\end{eqnarray}
where $\varepsilon_1$ and $\varepsilon_2$ are arbitrary numbers,
$\omega = e^{2\pi i/3}$, 
and $\overline{\omega} = \omega^2 = e^{4\pi i/3} (= -1 -\omega)$.
The above formulas are merely examples.
Quark kinetic coefficients and unitary matrices might take complicated forms
and contain tiny parameters intricately.
At any rate, a large mass hierarchy and mixing can originate from 
a tiny variance of the democratic form
whose every component has a common value.
In other words, {\it the hierarchical structure can be realized
in case that K${\ddot{a}}$hler metrics $K_{ij}^{(u)}$
and $K_{ij}^{(d)}$ acquire the VEVs of 
semi-democratic forms as
\begin{eqnarray}
\left\langle K_{ij}^{(u)} \right\rangle = \xi^{(u)} S_{ij} + O(\varepsilon_i),~~
\left\langle K_{ij}^{(d)} \right\rangle = \xi^{(d)} S_{ij} + O(\varepsilon_i, \lambda)
\label{<K>D}
\end{eqnarray}
with some constants $\xi^{(u)}$ and $\xi^{(d)}$,
after the breakdown of the family symmetry,
and the reception of tiny corrections.}
Here, $S_{ij}$ is the democratic matrix defined by
\begin{eqnarray}
S_{ij} \equiv
\left(
\begin{array}{ccc}
1 & 1 & 1 \\
1 & 1 & 1 \\
1 & 1 & 1
\end{array} 
\right).
\label{Democratic}
\end{eqnarray}
It is hard to derive semi-democratic forms (\ref{<K>D})
dynamically at a level of perturbation,
from U(3) invariant K\"{a}hler potential
$K = |\varPhi_i|^2 + \cdots$, as suggested by a model in Appendix B.
Here the ellipsis stands for higher-dimensional terms
which are sub-leading order ones.
We need a mechanism to realize semi-democratic forms
and small Yukawa couplings such as 
$\displaystyle{\left|y_1\right|^2 = O\left(10^{-10}\right)}$ 
and $\displaystyle{\left|y_2\right|^2 = O\left(10^{-9}\right)}$.

\subsubsection{S$_3$ case}

Based on an ${\rm S}_3$ invariant K\"{a}hler potential
containing the democratic form and Yukawa couplings
with the democratic form and small S$_3$ breaking ones,
it was pointed that the heavy top quark mass can be attributed to
a singular normalization of its kinetic term~\cite{KY}.
Sfermion masses were also studied using 
the ${\rm S}_3$ invariant K\"{a}hler potential~\cite{HKY}.

Let us re-examine a case with the S$_3$ symmetry
using our formulation.
Strictly speaking, the flavor group is ${\rm S}_3 \times {\rm S}_3 \times {\rm S}_3$,
and $q_{{\rm L}i}$, $u_{{\rm R}i}$ and $d_{{\rm R}i}$ are
transformed as 3-dimensional representations of the first, second and third S$_3$, respectively.
These 3-dimensional representations are reducible
and are decomposed into two irreducible ones
such as 1-dimensional ones and 2-dimensional ones.
In the presence of S$_3$ symmetry,
the Yukawa couplings are written by
\begin{eqnarray}
\left(y_1\right)_{ij} = y_1^{\rm F} S_{ij} + \varDelta y_1 T_{ij}^{(u)},~~
\left(y_2\right)_{ij} = y_2^{\rm F} S_{ij} + \varDelta y_2 T_{ij}^{(d)},
\label{y-S3}
\end{eqnarray}
where $y_a^{\rm F}$ and $\varDelta y_a$ ($a=1,2$) are complex numbers,
and $T_{ij}^{(u)}$ and $T_{ij}^{(d)}$ are complex matrices 
(whose components take values of at most $O(1)$) that originate from
S$_3$ breaking effects.
We cannot derive realistic quark masses without $T_{ij}^{(u)}$ and $T_{ij}^{(d)}$.
We assume that $\displaystyle{\left|y_a^{\rm F}\right| = O(1)}$ according to Dirac's naturalness.
Here, Dirac's naturalness means
that {\it the magnitude of dimensionless parameters on terms allowed by symmetries
should be $O(1)$ in a fundamental theory}.
In contrast, we suppose that $\displaystyle{\left|\varDelta y_a\right| \ll \left|y_a^{\rm F}\right|}$
from a conjecture that the S$_3$ breaking terms stem from
non-renormalizable interactions suppressed by a power of $M_{\rm BSM}$.

In the following, we examine whether 
magnitudes of components in $k_{ij}^{(u)}$ and $k_{ij}^{(d)}$
can be at most $O(1)$ or not under the above assumptions, i.e.,
$\displaystyle{\left|y_a^{\rm F}\right| = O(1)}$
and $\displaystyle{\left|\varDelta y_a\right| \ll \left|y_a^{\rm F}\right|}$.
In other words, $k_{ij}^{(u)}$ and $k_{ij}^{(d)}$ are, in general, written by
\begin{eqnarray}
k_{ij}^{(u)} = k_1^{(u)} \delta_{ij} + k_2^{(u)} S_{ij} + k_3^{(u)} Z_{ij}^{(u)},~~
k_{ij}^{(d)} = k_1^{(d)} \delta_{ij} + k_2^{(d)} S_{ij} + k_3^{(d)} Z_{ij}^{(d)},
\label{k-S3}
\end{eqnarray}
where $k_b^{(u)}$ and $k_b^{(d)}$ ($b = 1, 2, 3$) are real numbers,
and $Z_{ij}^{(u)}$ and $Z_{ij}^{(d)}$ are hermitian matrices 
(whose components take values of at most $O(1)$) that represent
S$_3$ breaking effects.
Then, can magnitudes of $k_b^{(u)}$ and $k_b^{(d)}$ be
at most $O(1)$ or not?

By inserting the first relation of (\ref{y-S3}) into (\ref{ku-GF}),
the following relation is derived,
\begin{eqnarray}
&~& k_{ij}^{(u)} = \left|y_1^{\rm F}\right|^2 \left(S{W^{(u)}}^{\dagger} 
\left(y_{\rm diag}^{(u) -1}\right)^2 W^{(u)} S\right)_{ij}
+ \left|\varDelta y_1\right|^2 \left({T^{(u)}}^{\dagger}{W^{(u)}}^{\dagger} 
\left(y_{\rm diag}^{(u) -1}\right)^2 W^{(u)} T^{(u)}\right)_{ij}
\nonumber\\
&~& ~~~~~~~~~~~~ 
+ \left(y_1^{\rm F}\right)^* \varDelta y_1 \left(S{W^{(u)}}^{\dagger} 
\left(y_{\rm diag}^{(u) -1}\right)^2 W^{(u)} T^{(u)}\right)_{ij} + {\rm h.c.}.
\label{ku-S3}
\end{eqnarray}
Using the formula $SXS = (\sum_{i, j=1}^3 X_{ij}) S$,
we find that the following condition
should be fulfilled,
\begin{eqnarray}
\sum_{i, j=1}^3 \left({W^{(u)}}^{\dagger} 
\left(y_{\rm diag}^{(u) -1}\right)^2 W^{(u)}\right)_{ij} = O(1),
\label{cond-S3}
\end{eqnarray}
in order to make the magnitudes of first term in (\ref{ku-S3}) to be at most $O(1)$.
For simplicity, let us take an ansatz of $W^{(u)}$ such as
\begin{eqnarray}
W^{(u)} \equiv
\left(
\begin{array}{ccc}
w_{11}^{(u)} & w_{12}^{(u)} & -w_{11}^{(u)} - w_{12}^{(u)} \\
w_{21}^{(u)} & w_{22}^{(u)} & -w_{21}^{(u)} - w_{22}^{(u)} \\
w_{31}^{(u)} & w_{32}^{(u)} & w_{33}^{(u)}
\end{array} 
\right),
\label{Wu-S3}
\end{eqnarray}
where $w_{ij}^{(u)}$ are complex numbers of at most $O(1)$.
Then, we obtain the relation:
\begin{eqnarray}
&~& k_{ij}^{(u)} \doteqdot \left|y_1^{\rm F}\right|^2 
\left|w_{31}^{(u)} + w_{32}^{(u)} + w_{33}^{(u)}\right|^2 S_{ij}
\nonumber \\
&~& ~~~~~~~~~~~~
+ \left|\varDelta y_1\right|^2 \left({T^{(u)}}^{\dagger}{W^{(u)}}^{\dagger} 
\left(y_{\rm diag}^{(u) -1}\right)^2 W^{(u)} T^{(u)}\right)_{ij} + O(|\varDelta y_1|).
\label{ku-S3-Wu}
\end{eqnarray}
If $\displaystyle{\left|\varDelta y_1\right|^2 = O\left(10^{-10}\right)}$,
the magnitude of every component in the second term of (\ref{ku-S3-Wu})
can also be at most $O(1)$, and $k_{ij}^{(u)}$ can take the form given by
the first relation of (\ref{k-S3}) with $\displaystyle{\left|y_1^{\rm F}\right| = O(1)}$.

In the same way, 
when we take an ansatz of $W^{(d)}$ such as
\begin{eqnarray}
W^{(d)} \equiv
\left(
\begin{array}{ccc}
w_{11}^{(d)} & w_{12}^{(d)} & -w_{11}^{(d)} - w_{12}^{(d)} \\
w_{21}^{(d)} & w_{22}^{(d)} & -w_{21}^{(d)} - w_{22}^{(d)} \\
w_{31}^{(d)} & w_{32}^{(d)} & w_{33}^{(d)}
\end{array} 
\right),
\label{Wd-S3}
\end{eqnarray}
we obtain the relation:
\begin{eqnarray}
&~& k_{ij}^{(d)} \doteqdot \left|y_2^{\rm F}\right|^2 \times 2.1 \times 10^{3} \times
\left|w_{31}^{(d)} + w_{32}^{(d)} + w_{33}^{(d)}\right|^2 S_{ij}
\nonumber \\
&~& ~~~~~~~~~~~~
+ \left|\varDelta y_2\right|^2 \left({T^{(d)}}^{\dagger}{W^{(d)}}^{\dagger} 
\left(y_{\rm diag}^{(d) -1}\right)^2 W^{(d)} T^{(d)}\right)_{ij} + O(|\varDelta y_2|),
\label{kd-S3-Wd}
\end{eqnarray}
where $w_{ij}^{(d)}$ are complex numbers of at most $O(1)$.
From $W^{(u)} = V_{\rm KM} W^{(d)}$,
we obtain $\displaystyle{\left|w_{31}^{(d)} + w_{32}^{(d)} + w_{33}^{(d)}\right|^2 = O(1)}$.
Hence, we need $\displaystyle{\left|y_2^{\rm F}\right|^2 = O\left(10^{-3}\right)}$
and $\displaystyle{\left|\varDelta y_2\right|^2 = O\left(10^{-9}\right)}$
in order to make 
the magnitude of every component in the first and second terms of (\ref{kd-S3-Wd})
to be at most $O(1)$, respectively.

\subsection{Lepton sector}

We study the lepton sector in the SM.
In the absence of Majorana masses of right-handed neutrino singlets,
the same argument as the quarks holds in the replacement of fields and couplings.
Here, we consider the case with large Majorana masses
and a flavor symmetry in a theory beyond the SM.

The lepton sector is described by the Lagrangian densities:
\begin{eqnarray}
&~& \mathscr{L}_{\rm kinetic}^{\rm lepton} = \overline{l}_{{\rm L}i} i \SlashD l_{{\rm L}i}
+ \overline{e}_{{\rm R}i} i \SlashD e_{{\rm R}i}
+ \overline{\nu}_{{\rm R}i} i \SlashD \nu_{{\rm R}i}
- \frac{1}{2} M_{ij} {\nu}_{{\rm R}i}^{\rm t} C \nu_{{\rm R}j},
\label{L-kin-lepton}\\
&~& \mathscr{L}_{\rm Yukawa}^{\rm lepton} 
= - y_{ij}^{(e)} \overline{l}_{{\rm L}i} \phi e_{{\rm R}j}
- y_{ij}^{(\nu)} \overline{l}_{{\rm L}i} \tilde{\phi} \nu_{{\rm R}j} + {\rm h.c.},
\label{L-Y-lepton}
\end{eqnarray}
where $l_{{\rm L}i}$ are left-handed lepton doublets, 
$e_{{\rm R}i}$ and $\nu_{{\rm R}i}$ are right-handed 
electron- and neutrino-type lepton singlets,
$M_{ij}$ are Majorana masses, ${\nu}_{{\rm R}i}^{\rm t}$ 
is a transpose of ${\nu}_{{\rm R}i}$,
$C=i\gamma^2 \gamma^0$,
and $y_{ij}^{(e)}$ and $y_{ij}^{(\nu)}$ are Yukawa couplings.
The $y_{ij}^{(e)}$ and $y_{ij}^{(\nu)}$ are diagonalized 
as $V_{\rm L}^{(e)} y^{(e)} {V_{\rm R}^{(e)}}^{\dagger} = y_{\rm diag}^{(e)}$
and $V_{\rm L}^{(\nu)} y^{(\nu)} {V_{\rm R}^{(\nu)}}^{\dagger} = y_{\rm diag}^{(\nu)}$
by bi-unitary transformations,
and $M_{ij}$ is also diagonalized by $V_{\rm R}^{(\nu)}$ as
$\displaystyle{{V_{\rm R}^{(\nu)}}^* M {V_{\rm R}^{(\nu)}}^{\dagger} = M_{\rm diag} 
= {\rm diag}\left(M_1, M_2, M_3\right)}$
under the assumption that the flavor symmetry exists beyond the SM.
Lepton masses are obtained as
\begin{eqnarray}
&~& V_{\rm L}^{(e)} y^{(e)} {V_{\rm R}^{(e)}}^{\dagger} \frac{v}{\sqrt{2}} 
= y_{\rm diag}^{(e)} \frac{v}{\sqrt{2}} 
= M_{\rm diag}^{(e)} = {\rm diag}\left(m_e, m_{\mu}, m_{\tau}\right),
\label{Me-diag}\\
&~& V_{\rm L}^{(\nu)} y^{(\nu)} M^{-1} {y^{(\nu)}}^{\rm t} {V_{\rm L}^{(\nu)}}^{\rm t} \frac{v^2}{2} 
= M_{\rm diag}^{(\nu)} = {\rm diag}\left(m_{\nu_1}, m_{\nu_{2}}, m_{\nu_{3}}\right),
\label{Mnu-diag}
\end{eqnarray}
where $V_{\rm L}^{(e)}$, $V_{\rm L}^{(\nu)}$, and $V_{\rm R}^{(e)}$  
are unitary matrices
and $m_e$, $m_{\mu}$, and $m_{\tau}$ are masses of electron, muon, and tauon, respectively,
and the seesaw mechanism is used
to obtain tiny neutrino masses $m_{\nu_1}$, $m_{\nu_{2}}$, and $m_{\nu_{3}}$~\cite{M,Y,GRS}. 
The lepton Yukawa couplings are expressed by
\begin{eqnarray}
y^{(e)} = {V_{\rm L}^{(e)}}^{\dagger} y_{\rm diag}^{(e)} V_{\rm R}^{(e)},~~
y^{(\nu)} = {V_{\rm L}^{(\nu)}}^{\dagger} y_{\rm diag}^{(\nu)} V_{\rm R}^{(\nu)}
= {V_{\rm L}^{(e)}}^{\dagger} V_{\rm MNS} y_{\rm diag}^{(\nu)} V_{\rm R}^{(\nu)},
\label{ynu}
\end{eqnarray}
using $V_{\rm L}^{(e)}$, $V_{\rm R}^{(e)}$, $V_{\rm R}^{(\nu)}$,
$y_{\rm diag}^{(e)}$, $y_{\rm diag}^{(\nu)}$,
and the Maki-Nakagawa-Sakata matrix 
$V_{\rm MNS} \equiv V_{\rm L}^{(e)} {V_{\rm L}^{(\nu)}}^{\dagger}$.

From (\ref{Me-diag}) and experimental values of charged lepton masses,
the magnitude of $y_{\rm diag}^{(e)}$ is roughly estimated at the weak scale as
\begin{eqnarray}
&~& y_{\rm diag}^{(e)} 
\doteqdot {\rm diag}\left(2.9 \times 10^{-6},~ 6.1 \times 10^{-4},~ 1.0 \times 10^{-2}\right).
\label{ye-diag-value}
\end{eqnarray}
We find that there is a hierarchy among charged lepton Yukawa couplings.

Using field variables $l'_{\rm L}$, $e'_{\rm R}$ and $\nu'_{\rm R}$ 
defined by
\begin{eqnarray}
&~& l'_{\rm L} \equiv U_l^{\dagger} J_l^{-1} V_l^{\dagger} l_{\rm L},~~
\left(\overline{l}'_{\rm L} \equiv \overline{l}_{\rm L} V_l J_l^{-1} U_l\right)
\label{l'}\\
&~& e'_{\rm R} \equiv y_3^{-1} U_l^{\dagger} J_l V_l^{\dagger} y^{(e)} e_{\rm R}
= y_3^{-1} U_l^{\dagger} J_l V_l^{\dagger}
{V_{\rm L}^{(e)}}^{\dagger} y_{\rm diag}^{(e)} V_{\rm R}^{(e)} e_{\rm R},~~
\label{e'}\\
&~& \nu'_{\rm R} \equiv y_4^{-1} U_l^{\dagger} J_l V_l^{\dagger} y^{(\nu)} \nu_{\rm R}
= y_4^{-1} U_l^{\dagger} J_l V_l^{\dagger}
{V_{\rm L}^{(\nu)}}^{\dagger} y_{\rm diag}^{(\nu)} V_{\rm R}^{(\nu)} \nu_{\rm R}
\nonumber \\
&~& ~~~~~ = y_4^{-1} U_l^{\dagger} J_l V_l^{\dagger}
{V_{\rm L}^{(e)}}^{\dagger} V_{\rm MNS} y_{\rm diag}^{(\nu)} V_{\rm R}^{(\nu)} \nu_{\rm R},
\label{nu'}
\end{eqnarray}
the Lagrangian densities are rewritten as
\begin{eqnarray}
&~& \mathscr{L}_{\rm kinetic}^{\rm lepton} = k_{ij}^{(l)} \overline{l}'_{{\rm L}i} i \SlashD l'_{{\rm L}j}
+ k_{ij}^{(e)} \overline{e}'_{{\rm R}i} i \SlashD e'_{{\rm R}j}
+ k_{ij}^{(\nu)} \overline{\nu}'_{{\rm R}i} i \SlashD \nu'_{{\rm R}j}
- \frac{1}{2} M_{ij}^{(\nu)} {\nu'}_{{\rm R}i}^{\rm t} C \nu'_{{\rm R}j},
\label{L-kin-lepton-prime}\\
&~& \mathscr{L}_{\rm Yukawa}^{\rm lepton} 
= - \left(y_3\right)_{ij} \overline{l}'_{{\rm L}i} {\phi} e'_{{\rm R}j}
- \left(y_4\right)_{ij} \overline{l}'_{{\rm L}i} \tilde{\phi} \nu'_{{\rm R}j} + {\rm h.c.},
\label{L-Y-lepton-prime}
\end{eqnarray}
where $V_l$ and $U_l$ are unitary matrices, $J_l$ is a real diagonal matrix,
$J_l^{-1}$ is the inverse matrix of $J_l$,
$\displaystyle{\left(y_3\right)_{ij}}$ and $\displaystyle{\left(y_4\right)_{ij}}$ 
are lepton Yukawa couplings in the unitary bases of flavor symmetry
and $k_{ij}^{(l)}$, $k_{ij}^{(e)}$, $k_{ij}^{(\nu)}$, and $M_{ij}^{(\nu)}$ are given by
\begin{eqnarray}
&~& k_{ij}^{(l)} = \left(U_l^{\dagger} (J_{l})^{2} U_l\right)_{ij},
\label{Kl}\\
&~& k_{ij}^{(e)} = \left(y_3^{\dagger}{W^{(e)}}^{\dagger} 
\left(y_{\rm diag}^{(e) -1}\right)^2 W^{(e)} y_3\right)_{ij},
\label{Ke}\\
&~& k_{ij}^{(\nu)} = \left(y_4^{\dagger} {W^{(\nu)}}^{\dagger} 
\left(y_{\rm diag}^{(\nu) -1}\right)^2 W^{(\nu)}y_4\right)_{ij}
= \left(y_4^{\dagger}{W^{(e)}}^{\dagger} V_{\rm MNS} \left(y_{\rm diag}^{(\nu) -1}\right)^2 
V_{\rm MNS}^{\dagger} W^{(e)} y_4\right)_{ij},
\label{Knu}\\
&~& M_{ij}^{(\nu)} = \left(y_4^{\rm t} {W^{(\nu)}}^{\rm t} 
y_{\rm diag}^{(\nu) -1} M_{\rm diag} y_{\rm diag}^{(\nu) -1} W^{(\nu)}y_4\right)_{ij}
\nonumber \\
&~& ~~~~~~~~~ = \left(y_4^{\rm t}{W^{(e)}}^{\rm t} V_{\rm MNS}^{*} y_{\rm diag}^{(\nu) -1} M_{\rm diag} 
y_{\rm diag}^{(\nu) -1} V_{\rm MNS}^{\dagger} W^{(e)} y_4\right)_{ij}.
\label{Mnu}
\end{eqnarray}
Here, $W^{(e)} = V_{\rm L}^{(e)} V_l J_{l}^{-1} U_l$ and 
$W^{(\nu)} = V_{\rm L}^{(\nu)} V_l J_{l}^{-1} U_l$.
From (\ref{ye-diag-value}),
$\left(y_{\rm diag}^{(e) -1}\right)^2$ is roughly estimated 
at the weak scale as
\begin{eqnarray}
&~& \left(y_{\rm diag}^{(e) -1}\right)^2 
\doteqdot {\rm diag}\left(1.2 \times 10^{11},~ 2.7 \times 10^{6},~ 1.0 \times 10^{4}\right)
\nonumber \\
&~& ~~~~~~~~~~~~~~~~~
= 1.2 \times 10^{11} \times {\rm diag}\left(1,~ 2.2 \times 10^{-5},~ 8.3 \times 10^{-8}\right).
\label{ye^(-2)-diag-value}
\end{eqnarray}

When a theory of lepton beyond the SM can be described by 
\begin{eqnarray}
&~& \mathscr{L}_{\rm BSM}^{\rm lepton} 
= K_{ij}^{(l)} \overline{l}'_{{\rm L}i} i \SlashD l'_{{\rm L}j}
+ K_{ij}^{(e)} \overline{e}'_{{\rm R}i} i \SlashD e'_{{\rm R}j}
+ K_{ij}^{(\nu)} \overline{\nu}'_{{\rm R}i} i \SlashD \nu'_{{\rm R}j}
- \frac{1}{2} \hat{M}_{ij}^{(\nu)} {\nu'}_{{\rm R}i}^{\rm t} C \nu'_{{\rm R}j}
\nonumber \\
&~&  ~~~~~~~~~~~~~~~~~~ - \left(Y_3\right)_{ij} \overline{l}'_{{\rm L}i} {\phi} e'_{{\rm R}j}
- \left(Y_4\right)_{ij} \overline{l}'_{{\rm L}i} \tilde{\phi} \nu'_{{\rm R}j} + {\rm h.c.},
\label{L-BSM-lepton}
\end{eqnarray}
we have the relations:
\begin{eqnarray}
&~& k_{ij}^{(l)} = \left\langle K_{ij}^{(l)} \right\rangle, ~~
k_{ij}^{(e)} = \left\langle K_{ij}^{(e)} \right\rangle,~~ 
k_{ij}^{(\nu)} = \left\langle K_{ij}^{(\nu)} \right\rangle,~~
{M}_{ij}^{(\nu)} = \left\langle \hat{M}_{ij}^{(\nu)} \right\rangle,~~
\label{<K>lepton}\\
&~& \left(y_3\right)_{ij} = \left\langle \left(Y_3\right)_{ij} \right\rangle,~~
\left(y_4\right)_{ij} = \left\langle \left(Y_4\right)_{ij} \right\rangle,
\label{<KY>lepton}
\end{eqnarray}
as the matching conditions at $M_{\rm BSM}$,
from (\ref{L-kin-lepton-prime}), (\ref{L-Y-lepton-prime})
and (\ref{L-BSM-lepton}).

In case that the U(3) family symmetry exists and 
$\displaystyle{\left|y_3\right|^2 = O\left(10^{-11}\right)}$,
the VEV of $K_{ij}^{(e)}$ can be the form
whose every component has an almost same magnitude of $O(1)$
and a mass hierarchy can originate from 
a tiny variance of the democratic form.
We need a mechanism to realize semi-democratic forms
and a small Yukawa coupling.
In case that S$_3$ flavor symmetry exists,
we find that a Yukawa coupling is written by
\begin{eqnarray}
\left(y_3\right)_{ij} = y_3^{\rm F} S_{ij} + \varDelta y_3 T_{ij}^{(e)},
\label{y-S3-lepton}
\end{eqnarray}
and it is compatible with the K\"{a}hler metric:
\begin{eqnarray}
k_{ij}^{(e)} = k_1^{(e)} \delta_{ij} + k_2^{(e)}  S_{ij} + k_3^{(e)} Z_{ij}^{(e)}
\label{k-S3-lepton}
\end{eqnarray}
with a suitable $W^{(e)}$.
Here, $y_3^{\rm F}$ and $\varDelta y_3$ are complex numbers
whose magnitudes are $\displaystyle{\left|y_3^{\rm F}\right|^2 = O(10^{-4})}$
and $\displaystyle{\left|\varDelta y_3\right|^2 = O\left(10^{-11}\right)}$,
$T_{ij}^{(e)}$ is a complex matrix whose components take values of at most $O(1)$,
$k_b^{(e)}$ ($b = 1, 2, 3$) are real numbers of at most $O(1)$,
and $Z_{ij}^{(e)}$ is a hermitian matrix 
whose components take values of at most $O(1)$.

\subsection{Top-down approach}

We have developed the strategy taking the SM as a starting point.
There are limitations on such a bottom-up approach.
It is desirable to combine use of the bottom-up
and top-down one.
Here, we propose a new procedure based on the top-down one, 
using knowledge and information obtained in the previous subsections.

First, we construct a theory with a flavor symmetry,
extract fermion parts from it
and write down a Lagrangian density as
\begin{eqnarray}
\hspace{-1cm}&~& \mathscr{L}_{\rm BSM}^{\rm fermion} 
= K_{ij}^{(q)} \overline{q}'_{{\rm L}i} i \SlashD q'_{{\rm L}j}
+ K_{ij}^{(u)} \overline{u}'_{{\rm R}i} i \SlashD u'_{{\rm R}j}
+ K_{ij}^{(d)} \overline{d}'_{{\rm R}i} i \SlashD d'_{{\rm R}j}
\nonumber \\
\hspace{-1cm}&~& ~~~~~~~~~~ + K_{ij}^{(l)} \overline{l}'_{{\rm L}i} i \SlashD l'_{{\rm L}j}
+ K_{ij}^{(e)} \overline{e}'_{{\rm R}i} i \SlashD e'_{{\rm R}j}
+ K_{ij}^{(\nu)} \overline{\nu}'_{{\rm R}i} i \SlashD \nu'_{{\rm R}j}
- \frac{1}{2} \hat{M}_{ij}^{(\nu)} {\nu'}_{{\rm R}i}^{\rm t} C \nu'_{{\rm R}j}
\nonumber \\
\hspace{-1cm}&~& ~~~~~~~~~~ - \left(Y_1\right)_{ij} \overline{q}'_{{\rm L}i} \tilde{\phi} u'_{{\rm R}j}
- \left(Y_2\right)_{ij} \overline{q}'_{{\rm L}i} \phi d'_{{\rm R}j} 
- \left(Y_3\right)_{ij} \overline{l}'_{{\rm L}i} \phi e'_{{\rm R}j} 
- \left(Y_4\right)_{ij} \overline{l}'_{{\rm L}i} \tilde{\phi} \nu'_{{\rm R}j}
+ {\rm h.c.},
\label{L-BSM-fermion-prime}
\end{eqnarray}
and obtain the VEVs of flavons from the minimum of a scalar potential.
Then, we calculate $\left\langle K_{ij}^{(q)} \right\rangle$, 
$\left\langle K_{ij}^{(u)} \right\rangle$, $\left\langle K_{ij}^{(d)} \right\rangle$, 
$\left\langle K_{ij}^{(l)} \right\rangle$, $\left\langle K_{ij}^{(e)} \right\rangle$,
$\left\langle K_{ij}^{(\nu)} \right\rangle$, 
$\left\langle \hat{M}_{ij}^{(\nu)} \right\rangle$,
$\left\langle\left(Y_1\right)_{ij}\right\rangle$,
$\left\langle\left(Y_2\right)_{ij}\right\rangle$, 
$\left\langle\left(Y_3\right)_{ij}\right\rangle$, and $\left\langle\left(Y_4\right)_{ij}\right\rangle$.
If the SUSY or its remnant exists,
$\tilde{\phi}$ and $\phi$ should be treated as independent fields.

Second, we diagonalize $\left\langle K_{ij}^{(q)} \right\rangle$
and $\left\langle K_{ij}^{(l)} \right\rangle$
by unitary transformations as
\begin{eqnarray}
\left(\tilde{U}_{q}\right)_{ii'} \left\langle K_{i'j'}^{(q)} \right\rangle 
\left(\tilde{U}_{q}^{\dagger}\right)_{j'j} = \left(\tilde{J}_q\right)_{ij}^2,~~
\left(\tilde{U}_{l}\right)_{ii'} \left\langle K_{i'j'}^{(l)} \right\rangle 
\left(\tilde{U}_{l}^{\dagger}\right)_{j'j} = \left(\tilde{J}_l\right)_{ij}^2,
\label{UKUdagger}
\end{eqnarray}
where $\tilde{U}_q$ and $\tilde{U}_l$ are unitary matrices
and $\tilde{J}_q$ and $\tilde{J}_l$ are real diagonal matrices.
These matrices are counterparts of $U_q$, $U_l$, $J_q$, and $J_l$,
and they should equate each other if experimental data 
on flavor physics are completely explained by them.

Third, we change $\left\langle K_{ij}^{(u)} \right\rangle$, $\left\langle K_{ij}^{(d)} \right\rangle$, 
$\left\langle K_{ij}^{(e)} \right\rangle$, and $\left\langle K_{ij}^{(\nu)} \right\rangle$
into the following ones,
\begin{eqnarray}
&~& \left\langle \tilde{K}_{ij}^{(u)} \right\rangle \equiv \left(\tilde{J}_{q}\right)_{ii'} 
\left(\tilde{U}_{q}\right)_{i'i''} \left\langle\left(Y_1^{\dagger -1}\right)_{i''i'''}\right\rangle
\left\langle K_{i'''j'''}^{(u)} \right\rangle \left\langle\left(Y_1^{-1}\right)_{j'''j''}\right\rangle
\left(\tilde{U}_{q}^{\dagger}\right)_{j''j'} \left(\tilde{J}_{q}\right)_{j'j},
\label{tildeKu}\\
&~& \left\langle \tilde{K}_{ij}^{(d)} \right\rangle \equiv \left(\tilde{J}_{q}\right)_{ii'} 
\left(\tilde{U}_{q}\right)_{i'i''} \left\langle\left(Y_2^{\dagger -1}\right)_{i''i'''}\right\rangle
\left\langle K_{i'''j'''}^{(d)} \right\rangle \left\langle\left(Y_2^{-1}\right)_{j'''j''}\right\rangle
\left(\tilde{U}_{q}^{\dagger}\right)_{j''j'} \left(\tilde{J}_{q}\right)_{j'j},
\label{tildeKd}\\
&~& \left\langle \tilde{K}_{ij}^{(e)} \right\rangle \equiv \left(\tilde{J}_{l}\right)_{ii'} 
\left(\tilde{U}_{l}\right)_{i'i''} \left\langle\left(Y_3^{\dagger -1}\right)_{i''i'''}\right\rangle
\left\langle K_{i'''j'''}^{(e)} \right\rangle \left\langle\left(Y_3^{-1}\right)_{j'''j''}\right\rangle
\left(\tilde{U}_{l}^{\dagger}\right)_{j''j'} \left(\tilde{J}_{l}\right)_{j'j},
\label{tildeKe}\\
&~& \left\langle \tilde{K}_{ij}^{(\nu)} \right\rangle \equiv \left(\tilde{J}_{l}\right)_{ii'} 
\left(\tilde{U}_{l}\right)_{i'i''} \left\langle\left(Y_4^{\dagger -1}\right)_{i''i'''}\right\rangle
\left\langle K_{i'''j'''}^{(\nu)} \right\rangle \left\langle\left(Y_4^{-1}\right)_{j'''j''}\right\rangle
\left(\tilde{U}_{l}^{\dagger}\right)_{j''j'} \left(\tilde{J}_{l}\right)_{j'j},
\label{tildeKnu}
\end{eqnarray}
using $\tilde{U}_q$, $\tilde{U}_l$, 
$\tilde{J}_q$, $\tilde{J}_l$,
the inverse matrices $\left\langle Y_a^{-1}\right\rangle$ of 
$\left\langle Y_a \right\rangle$ ($a=1, 2, 3, 4$)
and their hermitian conjugations.

Fourth, we diagonalize $\left\langle \tilde{K}_{ij}^{(u)} \right\rangle$, 
$\left\langle \tilde{K}_{ij}^{(d)} \right\rangle$, 
$\left\langle \tilde{K}_{ij}^{(e)} \right\rangle$, and $\left\langle \tilde{K}_{ij}^{(\nu)} \right\rangle$
by unitary transformations as
\begin{eqnarray}
&~& \left(\tilde{V}_{\rm L}^{(u)}\right)_{ii'} \left\langle \tilde{K}_{i'j'}^{(u)} \right\rangle
\left(\tilde{V}_{\rm L}^{(u)\dagger}\right)_{j'j} = \left(\tilde{k}_{\rm diag}^{(u)}\right)_{ij},~~
\left(\tilde{V}_{\rm L}^{(d)}\right)_{ii'} \left\langle \tilde{K}_{i'j'}^{(d)} \right\rangle
\left(\tilde{V}_{\rm L}^{(d)\dagger}\right)_{j'j} = \left(\tilde{k}_{\rm diag}^{(d)}\right)_{ij},
\label{tildeKquark-diag}\\
&~& \left(\tilde{V}_{\rm L}^{(e)}\right)_{ii'} \left\langle \tilde{K}_{i'j'}^{(e)} \right\rangle
\left(\tilde{V}_{\rm L}^{(e)\dagger}\right)_{j'j} = \left(\tilde{k}_{\rm diag}^{(e)}\right)_{ij},~~
\left(\tilde{V}_{\rm L}^{(\nu)}\right)_{ii'} \left\langle \tilde{K}_{i'j'}^{(\nu)} \right\rangle
\left(\tilde{V}_{\rm L}^{(\nu)\dagger}\right)_{j'j} = \left(\tilde{k}_{\rm diag}^{(\nu)}\right)_{ij}.
\label{tildeKlepton-diag}
\end{eqnarray}

Last, we examine whether the following relations hold or not,
\begin{eqnarray}
\hspace{-0.5cm}&~& \left(\tilde{k}_{\rm diag}^{(u)}\right)_{ij} 
= \left(y_{\rm diag}^{(u) -1}\right)^2_{ij},~~
 \left(\tilde{k}_{\rm diag}^{(d)}\right)_{ij} = \left(y_{\rm diag}^{(d) -1}\right)^2_{ij},~~
 \left(\tilde{k}_{\rm diag}^{(e)}\right)_{ij} = \left(y_{\rm diag}^{(e) -1}\right)^2_{ij},~~
\label{tildeK-check}\\
\hspace{-0.5cm}&~& \tilde{V}_{\rm L}^{(u)} \tilde{V}_{\rm L}^{(d)\dagger} = V_{\rm KM},~~
\tilde{V}_{\rm L}^{(e)} \tilde{V}_{\rm L}^{(\nu)\dagger} = V_{\rm MNS}.
\label{tildeV-check}
\end{eqnarray}
Note that we need to diagonalize six hermitian matrices in total
by unitary transformations in our procedure.
As explained in Appendix C, 
we need ten hermitian matrices in total
by unitary transformations, 
using an ordinary procedure.

As was described previously,
we should consider renormalization effects when we match
theoretical predictions to experimental data.
We also need some modifications in the presence of a mixing with extra particles,
in the case with a large flavor symmetry
and/or many matter fields.

\subsection{Unification}

We discuss whether realistic mass hierarchies and flavor mixing are realized or nor,
based on a grand unification and a family unification.

First, we consider a model based on ${\rm SU}(5) \times {\rm S}_3 \times {\rm S}_3$
where ${\rm SU}(5)$ is the GUT group and ${\rm S}_3 \times {\rm S}_3$ is the flavor group.
We assume that these symmetries are broken down to the SM one ${\rm G}_{\rm SM}$
at the GUT scale $M_{\rm U}$.
Matter fields $l'_{{\rm L}i}$ and $(d'_{{\rm R}i})^c$ belong to 
${\psi'}_i^{(\overline{\bm{5}})}$ in the representation
$(\overline{\bm{5}}, \bm{3}, \bm{1})$ 
and $q'_{{\rm L}i}$, $(u'_{{\rm R}i})^c$ and $(e'_{{\rm R}i})^c$
belong to ${\psi'}_i^{({\bm{10}})}$ in $(\bm{10}, \bm{1}, \bm{3})$,
where $\bm{3}$ is a 3-dimensional reducible representation of S$_3$.
The Lagrangian density of matter fields (except for neutrino singlets) is given by
\begin{eqnarray}
&~& \mathscr{L}_{\rm SU(5)}^{\rm fermion}
= K_{ij}^{(\psi^{(\overline{\bm{5}})})} 
{\overline{\psi}'}_i^{(\overline{\bm{5}})} i \SlashD {\psi'}_j^{(\overline{\bm{5}})}
 + K_{ij}^{(\psi^{(\bm{10})})} {\overline{\psi}'}_i^{(\bm{10})} i \SlashD {\psi'}_j^{(\bm{10})}
\nonumber \\
&~& ~~~~~~~~~~~~~~~~~~~~~~
 - \left(Y_1^{\rm U}\right)_{ij} 
{{\psi'}_i^{({\bm{10}})}}^{\rm t} C {\psi'}_j^{(\overline{\bm{5}})} \phi^{(\overline{\bm{5}})}
- \left(Y_2^{\rm U}\right)_{ij}
{{\psi'}_i^{({\bm{10}})}}^{\rm t} C {\psi'}_j^{({\bm{10}})} \phi^{\bm{(5)}} + {\rm h.c.},
\label{KY-SU5}
\end{eqnarray}
where $\left(Y_1^{\rm U}\right)_{ij}$ and $\left(Y_2^{\rm U}\right)_{ij}$ 
are Yukawa couplings,
and $\phi^{(\overline{\bm{5}})}$ and $\phi^{\bm{(5)}}$ are scalar fields
in $(\overline{\bm{5}}, \bm{1}, \bm{1})$ and $(\bm{5}, \bm{1}, \bm{1})$, respectively.
If $K_{ij}^{(\psi^{(\overline{\bm{5}})})}$, $K_{ij}^{(\psi^{(\bm{10})})}$,
$\left(Y_1^{\rm U}\right)_{ij}$, and $\left(Y_2^{\rm U}\right)_{ij}$ are ${\rm SU(5)}$ singlets,
we have the relations:
\begin{eqnarray}
&~& \left\langle K_{ij}^{(\psi^{(\overline{\bm{5}})})} \right\rangle
= \left\langle K_{ij}^{(l)} \right\rangle =  \left\langle K_{ij}^{(d)} \right\rangle,~~
\left\langle K_{ij}^{(\psi^{(\bm{10})})} \right\rangle
= \left\langle K_{ij}^{(q)} \right\rangle =  \left\langle K_{ij}^{(u)} \right\rangle 
= \left\langle K_{ij}^{(e)} \right\rangle,~~
\label{K-SU5}\\
&~& \left\langle \left(Y_1^{\rm U}\right)_{ij} \right\rangle
= \left\langle \left(y_2\right)_{ij} \right\rangle
= \left\langle \left(y_3\right)_{ji} \right\rangle,~~
 \left\langle \left(Y_2^{\rm U}\right)_{ij} \right\rangle
= \left\langle \left(y_1\right)_{ij} \right\rangle,
\label{Y-SU5}
\end{eqnarray}
at $M_{\rm U}$.
From (\ref{K-SU5}) and (\ref{Y-SU5}),
we derive a usual GUT relation among down-type quark and charged lepton 
Yukawa couplings:
\begin{eqnarray}
\left(y^{(d)}\right)_{ij} = \left(y^{(e)}\right)_{ji}.
\label{Y-SU5-relation}
\end{eqnarray}
In case that $\left(Y_1^{\rm U}\right)_{ij}$ and $\left(Y_2^{\rm U}\right)_{ij}$
contain ${\rm SU}(5)$ non-singlet parts,
realistic mass hierarchies and mixing can be realized
with suitable VEVs of non-singlet parts.

Next, we consider a model based on ${\rm SO}(10) \times {\rm S}_3$.
Matter fields $l'_{{\rm L}i}$, $(d'_{{\rm R}i})^c$,
$q'_{{\rm L}i}$, $(u'_{{\rm R}i})^c$, $(e'_{{\rm R}i})^c$, and $(\nu'_{{\rm R}i})^c$
belong to ${\psi'}_i^{(\bm{16})}$ in $(\bm{16}, \bm{3})$.
The matter sector is described by
\begin{eqnarray}
\mathscr{L}_{\rm SO(10)}^{\rm fermion}
= K_{ij}^{(\psi^{(\bm{16})})} {\overline{\psi}'}_i^{(\bm{16})} i \SlashD {\psi'}_j^{(\bm{16})}
 - \left(\left(Y^{\rm U}\right)_{ij} {{\psi'}_i^{({\bm{16}})}}^{\rm t} C {\psi'}_j^{({\bm{16}})} \phi^{\bm{(10)}} 
+ {\rm h.c.}\right),
\label{KY-SO10}
\end{eqnarray}
where $\left(Y^{\rm U}\right)_{ij}$ is a Yukawa coupling 
and $\phi^{\bm{(10)}}$ is a scalar field in $(\bm{10}, \bm{1})$.
If $K_{ij}^{(\psi^{(\bm{16})})}$ and $\left(Y^{\rm U}\right)_{ij}$ are ${\rm SO}(10)$ singlets,
we have the relations:
\begin{eqnarray}
&~& \left\langle K_{ij}^{(\psi^{(\bm{16})})} \right\rangle =
\left\langle K_{ij}^{(l)} \right\rangle =  \left\langle K_{ij}^{(d)} \right\rangle
= \left\langle K_{ij}^{(q)} \right\rangle =  \left\langle K_{ij}^{(u)} \right\rangle 
= \left\langle K_{ij}^{(e)} \right\rangle
= \left\langle K_{ij}^{(\nu)} \right\rangle,
\label{K-SO10}\\
&~& \left\langle \left(Y^{\rm U}\right)_{ij} \right\rangle
= \left\langle \left(y_1\right)_{ij} \right\rangle
= \left\langle \left(y_2\right)_{ij} \right\rangle
= \left\langle \left(y_3\right)_{ij} \right\rangle
= \left\langle \left(y_4\right)_{ij} \right\rangle,
\label{Y-SO10}
\end{eqnarray}
at $M_{\rm U}$.
In this case, without extra matters and/or extra contributions, 
quark and lepton masses and flavor mixing cannot be explained.
In case that $\left(Y^{\rm U}\right)_{ij}$
contain ${\rm SO}(10)$ non-singlet parts,
we also need extra contributions if Dirac's naturalness is adopted.

Last, we consider the family unification based on a simple gauge group
${\rm G}_{\rm FU}$ whose maximal subgroup is ${\rm G}_{\rm U} \times {\rm G}_{\rm F}$.
Here, ${\rm G}_{\rm U}$ is a GUT group and ${\rm G}_{\rm F}$ is a family group.
We assume that a field $\varPsi$ with a vectorlike representation contains
three families of SM fermions $\psi_i^I =$ ($q'_{{\rm L}i}$, $(u'_{{\rm R}i})^c$, 
$(d'_{{\rm R}i})^c$, $l'_{{\rm L}i}$, $(e'_{{\rm R}i})^c$, $(\nu'_{{\rm R}i})^c$)
($I = q, u, d, l, e, \nu$) as its submultiplets.
After the breakdown of ${\rm G}_{\rm FU}$ into ${\rm G}_{\rm SM}$,
the kinetic term $K \overline{\varPsi} i \SlashD \varPsi$
changes into $\left\langle K_{ij}^{(I)} \right\rangle \overline{\psi}_i^I i \SlashD \psi_j^I$.
In this case, 
$\left\langle K_{ij}^{(I)} \right\rangle$ are not, in general, common 
and there is a possibility to explain fermion masses and flavor mixing.
However, it seems to be unnatural because we need a fine-tuning
on a realization of semi-democratic type of K\"{a}hler metrics
in order to generate fermion mass hierarchies, as explained in Appendix B.
Other problem in the family unification is 
that extra particles including mirror particles appear,
and it is solved in the family unification on orbifold~\cite{KKO,GKM,GK}
and special GUTs~\cite{Yamatsu,Yamatsu2}.

\section{Conclusions and discussions}

We have studied the origin of fermion mass hierarchy
and flavor mixing in the SM, using the bottom-up approach.
The approach is based on the assumptions
that the field variables in the SM are not necessarily the same
as those in a theory beyond the SM
and there is a flavor symmetry and flavons couple to matter fields
in the matter kinetic terms dominantly.
We have supposed field variables respecting a flavor symmetry
(unitary bases of a flavor symmetry)
and rewritten the Lagrangian density in the SM 
using such variables.
We have investigated the structure of terms violating 
the flavor symmetry,
and conjectured physics beyond the SM.
We have suggested that the hierarchical structure in the Yukawa interactions
of quarks and charged leptons
can originate from non-canonical matter kinetic terms,
in the presence of flavor symmetric Yukawa interactions
and a flavor symmetry can be hidden in the form of 
non-unitary bases in the SM.
We have proposed a variant top-down procedure,
using an insight and formulas obtained by our bottom-up approach.

In our approach, the problem of fermion masses and flavor mixing
is deeply related to not only the determination of Yukawa coupling matrices
but also the determination of matter kinetic terms
and the VEVs of K\"{a}hler metric $K_{ij}^{(I)}$.
If flavons couple to matter fields in the K\"{a}hler potential,
the VEVs of $K_{ij}^{(I)}$ strongly depend on the dynamics of flavor symmetry breaking
due to flavons.
In a grand unification with a flavor symmetry,
contributions of GUT group non-singlet parts in $K_{ij}^{(I)}$ 
can be essential to derive a realistic flavor structure.

We explain preceding works on the flavor physics based on matter kinetic terms,
other than \cite{KY,HKY}.
The problem of fermion mass hierarchies was investigated
in supergravity and superstring models
with non-canonical K\"{a}hler potential including dilaton and moduli fields~\cite{BD1,BD2}.
The Yukawa textures were obtained from non-canonical K\"{a}hler potential
in the extension of minimal SUSY SM with an anomalous horizontal
symmetry~\cite{BLR}.
In both works, a symmetry corresponding to a flavor symmetry
is an Abelian one and the structure of Yukawa couplings
resembles that derived from the Froggatt-Nielsen mechanism~\cite{FN}.
The effect of the K\"{a}hler potential on mixing matrices
was studied in a model independent way~\cite{EI}.
The flavor symmetry of kinetic terms was discussed
in a SUSY SM~\cite{Liu}.
The flavor problem was studied
through contributions of higher-dimensional operators 
in case with hierarchical fermion kinetic terms
originated from hierarchical fermion wave functions,
under the assumption that the energy scale of new physics is in the TeV range~\cite{DIU}.
In our setup, the scale $M_{\rm BSM}$ can also be constrained
by the suppression of flavor-changing transitions.

As fermion kinetic functions or K\"{a}hler metric $K_{ij}^{(I)}$
contain flavons in our approach,
they are regarded as counterparts of ``Yukawaons''
such that Yukawa couplings are not parameters but fields~\cite{Koide2}.

Our approach would be useful as a complementary one to explore
physics beyond the SM
and it would be worth studying flavor physics model-dependently
and/or independently by paying close attention to matter kinetic terms, 
because the structure of K\"{a}hler potential
can play a vital role as a key test of new physics.

\section*{Acknowledgments}
This work was supported in part by scientific grants 
from the Ministry of Education, Culture,
Sports, Science and Technology under Grant No.~17K05413.

\appendix

\section{Unitary and non-unitary bases}
\label{app-A}

We give an illustration
of a realization of U$(N)$ symmetry using unitary matrices and non-unitary ones
based on a polynomial:
\begin{eqnarray}
\mathscr{L} = \varPhi^{\dagger} K \varPhi + \varPhi^{\dagger} \varPhi,
\label{L-Phi}
\end{eqnarray}
where $\varPhi$ is an $N$-plet of U$(N)$, and $K$ 
is an $N \times N$ hermitian matrix.
We consider a case that $K$ depends on a set of fields $\{\varphi\}$, i.e.,
$K=K(\varphi, \varphi^{\dagger})$.
If $K$ changes into $K \to UKU^{\dagger}$ 
in accord with the U$(N)$ transformation $\varPhi \to U \varPhi$
with an arbitrary unitary matrix $U$,
$\mathscr{L}$ is invariant under the U$(N)$ transformation.
We call fields transformed by unitary matrices 
such as $\varPhi$ ``unitary bases''.

The U$(N)$ invariance can be spontaneously broken down
to a smaller one, 
after some $\varphi$ acquire the VEV $\langle \varphi \rangle$
and $\langle K \rangle (\equiv K(\langle \varphi \rangle, \langle \varphi^{\dagger} \rangle))$ 
takes a form that is not proportional to the identity matrix $I$.
The $\langle K \rangle$ is a hermitian matrix and it is written as
$\langle K \rangle = W^{\dagger} W$ with a general $N \times N$ complex matrix $W$.
By using a redefinition of field as $\tilde{\varPhi} \equiv W \varPhi$
and $\tilde{\varPhi}^{\dagger} \equiv \varPhi^{\dagger} W^{\dagger}$,
$\mathscr{L}$ is rewritten by
\begin{eqnarray}
\tilde{\mathscr{L}} =
\varPhi^{\dagger} \langle K \rangle \varPhi + \varPhi^{\dagger} \varPhi
= \varPhi^{\dagger} W^{\dagger} W \varPhi + \varPhi^{\dagger} \varPhi
= \tilde{\varPhi}^{\dagger} \tilde{\varPhi} 
+ \tilde{\varPhi}^{\dagger} (W^{\dagger})^{-1} W^{-1} \tilde{\varPhi}.
\label{L-tildevarPhi}
\end{eqnarray}
The previous U$(N)$ transformation is realized by
$\tilde{\varPhi} \to \tilde{U} \tilde{\varPhi}$
with $\tilde{U} = W U W^{-1}$.
Note that $\tilde{U}$ is not necessarily a unitary matrix
because $W$ is not a unitary matrix,
and the second term $\tilde{\varPhi}^{\dagger} (W^{\dagger})^{-1} W^{-1} \tilde{\varPhi}$
is invariant under $\tilde{\varPhi} \to \tilde{U} \tilde{\varPhi}$,
but the first one $\tilde{\varPhi}^{\dagger} \tilde{\varPhi}$ is not necessarily.
The transformation of unbroken subgroup H is realized by a unitary matrix.
We call fields transformed by non-unitary matrices 
such as $\tilde{\varPhi}$ ``non-unitary bases''.
The $\mathscr{L}$ and the final form of $\tilde{\mathscr{L}}$
can be regarded as counterparts of the Lagrangian density of matter sector
in a theory beyond the SM
and the Lagrangian density of matter sector in the SM, respectively.

\section{Non-canonical K\"{a}hler potential}
\label{app-B}

We consider a SUSY model with the flavor symmetry
${\rm SU}(3) \times {\rm C}_3$
(where ${\rm C}_3$ is the cyclic group of order 3)
and a non-minimal K\"{a}hler potential:
\begin{eqnarray}
&~& K = \left(1+ \frac{a_1}{\varLambda^2} \varphi_k^{\alpha} {\varphi_k^{\alpha}}^{\dagger}
+ \frac{a_2}{\varLambda^2} \sum_{\alpha} \varphi_k^{\alpha} 
\sum_{\beta} {\varphi_k^{\beta}}^{\dagger} \right) |\phi_i|^2
\nonumber \\
&~& ~~~~~~~
+  \left(\frac{a_3}{\varLambda^2} \varphi_i^{\alpha} {\varphi_j^{\alpha}}^{\dagger}
+ \frac{a_4}{\varLambda^2} \sum_{\alpha} \varphi_i^{\alpha} 
\sum_{\beta} {\varphi_j^{\beta}}^{\dagger} \right) \phi_i^{\dagger} \phi_j + \cdots,
\label{K-SUSY}
\end{eqnarray}
where $a_1$, $a_2$, $a_3$ and $a_4$ are parameters, $\varLambda$ is a high-energy scale,
and $\varphi_i^{\alpha}$ and $\phi_i$ are the scalar components of
flavon chiral supermultiplets and matter chiral supermultiplet, respectively.
The ellipsis stands for higher-dimensional terms with $\displaystyle{O\left(1/\varLambda^4\right)}$.
The family labels are denoted by $i$, $j$, and $k$, 
and $\varphi_i^{\alpha}$ and $\phi_i$ belong to triplets of ${\rm SU}(3)$.
The indices $\alpha$ and $\beta$ are labels of ${\rm C}_3$
and run from 1 to 3.
From (\ref{K-SUSY}),
the K\"{a}hler metric of matter fields is calculated as
\begin{eqnarray}
&~& K_{ij} = \frac{\partial^2K}{\partial\phi_i^{\dagger}\partial\phi_j}
=\left(1+ \frac{a_1}{\varLambda^2} \varphi_k^{\alpha} {\varphi_k^{\alpha}}^{\dagger}
+ \frac{a_2}{\varLambda^2} \sum_{\alpha} \varphi_k^{\alpha} 
\sum_{\beta} {\varphi_k^{\beta}}^{\dagger}\right) \delta_{ij}
\nonumber \\
&~& ~~~~~~~~~~~~~~~~~~~~~~~~~~~~~~~~
+ \frac{a_3}{\varLambda^2} \varphi_i^{\alpha} {\varphi_j^{\alpha}}^{\dagger}
+ \frac{a_4}{\varLambda^2} \sum_{\alpha} \varphi_i^{\alpha} 
\sum_{\beta} {\varphi_j^{\beta}}^{\dagger}  + \cdots.
\label{Kij-SUSY}
\end{eqnarray}
If $\varLambda$ is much bigger than the VEVs of $\varphi_i^{\alpha}$,
$\displaystyle{\left|\phi_i\right|^2}$ dominates in $K$
and the matter kinetic terms take almost canonical forms with 
$\displaystyle{\left\langle K_{ij} \right\rangle 
= \delta_{ij} + O\left(\left(\langle \varphi_i^{\alpha} \rangle/\varLambda\right)^2\right)}$.

To obtain a semi-democratic form, we need 
$\displaystyle{\left\langle \varphi_i^{\alpha} \right\rangle = O(\varLambda)}$.
In this case, other higher order terms can contribute the determination of 
$\displaystyle{\left\langle K_{ij} \right\rangle}$
and then the evaluation cannot be justified in a perturbation region.
Although we have such a problem,
we study a case with 
$\displaystyle{\left\langle \varphi_i^{\alpha} \right\rangle = O(\varLambda)}$
by taking the superpotential of flavons:
\begin{eqnarray}
W^{(\varphi)} = c_1 \varphi^3 + \frac{c_2}{\varLambda^3} \left(\varphi^3\right)^2,
\label{W-SUSY}
\end{eqnarray}
where $c_1$ and $c_2$ are parameters
and $\varphi^3 \equiv \varepsilon^{ijk} \varepsilon_{\alpha\beta\gamma}
\varphi_i^{\alpha} \varphi_j^{\beta} \varphi_k^{\gamma}$.
One of the SUSY preserving conditions is given by
\begin{eqnarray}
\frac{\partial W^{(\varphi)}}{\partial \varphi_i^{\alpha}} 
= 3 \varepsilon^{ijk} \varepsilon_{\alpha\beta\gamma}
\varphi_j^{\beta} \varphi_k^{\gamma}
\left(c_1 + \frac{2 c_2}{\varLambda^3}\varphi^3\right) = 0,
\label{dW-SUSY}
\end{eqnarray}
and there exist two kinds of vacuum solutions 
$\left\langle \varphi_i^{\alpha} \right\rangle = 0$ 
and $\left\langle \varphi_i^{\alpha} \right\rangle \ne 0$.\\
(a) Flavor symmetric vacuum with $\left\langle \varphi_i^{\alpha} \right\rangle = 0$\\
By inserting $\left\langle \varphi_i^{\alpha} \right\rangle = 0$ 
into (\ref{K-SUSY}) and (\ref{Kij-SUSY}),
we obtain the canonical one for matter fields, i.e.,
$\left\langle K_{ij} \right\rangle = \delta_{ij}$.\\
(b) Broken vacuum of flavor symmetry with $\left\langle \varphi_i^{\alpha} \right\rangle \ne 0$\\
From (\ref{dW-SUSY}),
we find a broken vacuum of flavor symmetry represented by
\begin{eqnarray}
\left\langle \varphi_i^{\alpha} \right\rangle = \left(\frac{-c_1}{2c_2}\right)^{1/3} 
\times \varLambda \delta_i^{\alpha}.
\label{vF}
\end{eqnarray}
Then, by inserting these VEVs into (\ref{Kij-SUSY}),
we obtain the VEV of $K_{ij}$:
\begin{eqnarray}
\left\langle K_{ij} \right\rangle = \eta
\left(
\begin{array}{ccc}
1 & 0 & 0 \\
0 & 1 & 0 \\
0 & 0 & 1
\end{array} 
\right)
+ \xi
\left(
\begin{array}{ccc}
1 & 1 & 1 \\
1 & 1 & 1 \\
1 & 1 & 1
\end{array} 
\right) + \cdots,
\label{<Kij>-SUSY}
\end{eqnarray}
where $\eta$ and $\xi$ are given by
\begin{eqnarray}
\eta = 1+\left(3a_1 + 3a_2 + a_3\right)\left(\frac{-c_1}{2c_2}\right)^{2/3},~~
\xi = a_4 \left(\frac{-c_1}{2c_2}\right)^{2/3},
\label{etaxi}
\end{eqnarray}
respectively.
From (\ref{<Kij>-SUSY}), $\left\langle K_{ij} \right\rangle$ can be
a semi-democratic one with suitable values of parameters,
but it seems to be unnatural with a fine-tuning among
parameters (including ones from higher order terms)
based on a perturbative analysis.
A K\"{a}hler potential from a non-perturbative effect
can play a crucial role to the derivation of semi-democratic types of kinetic terms.

\section{Ordinary top-down procedure}

For a purpose of reference, we explain an ordinary top-down procedure,
starting from $\mathscr{L}_{\rm BSM}^{\rm fermion}$ of (\ref{L-BSM-fermion-prime})
with the VEVs $\left\langle K_{ij}^{(q)} \right\rangle$, 
$\left\langle K_{ij}^{(u)} \right\rangle$, $\left\langle K_{ij}^{(d)} \right\rangle$, 
$\left\langle K_{ij}^{(l)} \right\rangle$, $\left\langle K_{ij}^{(e)} \right\rangle$,
$\left\langle K_{ij}^{(\nu)} \right\rangle$, 
$\left\langle \hat{M}_{ij}^{(\nu)} \right\rangle$,
$\left\langle\left(Y_1\right)_{ij}\right\rangle$,
$\left\langle\left(Y_2\right)_{ij}\right\rangle$, 
$\left\langle\left(Y_3\right)_{ij}\right\rangle$, and $\left\langle\left(Y_4\right)_{ij}\right\rangle$.

First, we diagonalize the K\"{a}hler metrics by unitary transformations as
\begin{eqnarray}
&~& \left(\tilde{U}_{q}\right)_{ii'} \left\langle K_{i'j'}^{(q)} \right\rangle 
\left(\tilde{U}_{q}^{\dagger}\right)_{j'j} = \left(\tilde{J}_q\right)_{ij}^2,~~
\left(\tilde{U}_{u}\right)_{ii'} \left\langle K_{i'j'}^{(u)} \right\rangle 
\left(\tilde{U}_{u}^{\dagger}\right)_{j'j} = \left(\tilde{J}_u\right)_{ij}^2,
\label{UKquUdagger}\\
&~& \left(\tilde{U}_{d}\right)_{ii'} \left\langle K_{i'j'}^{(d)} \right\rangle 
\left(\tilde{U}_{d}^{\dagger}\right)_{j'j} = \left(\tilde{J}_d\right)_{ij}^2,~~
\left(\tilde{U}_{l}\right)_{ii'} \left\langle K_{i'j'}^{(l)} \right\rangle 
\left(\tilde{U}_{l}^{\dagger}\right)_{j'j} = \left(\tilde{J}_l\right)_{ij}^2,
\label{UKdlUdagger}\\
&~& \left(\tilde{U}_{e}\right)_{ii'} \left\langle K_{i'j'}^{(e)} \right\rangle 
\left(\tilde{U}_{e}^{\dagger}\right)_{j'j} = \left(\tilde{J}_e\right)_{ij}^2,~~
\left(\tilde{U}_{\nu}\right)_{ii'} \left\langle K_{i'j'}^{(\nu)} \right\rangle 
\left(\tilde{U}_{\nu}^{\dagger}\right)_{j'j} = \left(\tilde{J}_{\nu}\right)_{ij}^2,
\label{UenuUdagger}
\end{eqnarray}
where $\tilde{U}_q$, $\tilde{U}_u$, $\tilde{U}_d$, $\tilde{U}_l$, $\tilde{U}_e$,
and $\tilde{U}_{\nu}$ are unitary matrices
and $\tilde{J}_q$, $\tilde{J}_u$, $\tilde{J}_d$, $\tilde{J}_l$, $\tilde{J}_e$, and $\tilde{J}_{\nu}$ 
are real diagonal matrices.

Second, we obtain the following Yukawa couplings from $\displaystyle{\left(Y_1\right)_{ij}}$, 
$\displaystyle{\left(Y_2\right)_{ij}}$, $\displaystyle{\left(Y_3\right)_{ij}}$
and $\displaystyle{\left(Y_4\right)_{ij}}$ such that
\begin{eqnarray}
&~& \tilde{y}_{ij}^{(u)} = \left(\tilde{J}_{q}^{-1}\right)_{ii'} 
\left(\tilde{U}_{q}\right)_{i'i''} \left\langle \left(Y_1\right)_{i''j''} \right\rangle
\left(\tilde{U}_{u}^{\dagger}\right)_{j''j'} \left(\tilde{J}_{u}^{-1}\right)_{j'j},
\label{tildeyu}\\
&~& \tilde{y}_{ij}^{(d)} = \left(\tilde{J}_{q}^{-1}\right)_{ii'} 
\left(\tilde{U}_{q}\right)_{i'i''} \left\langle \left(Y_2\right)_{i''j''} \right\rangle
\left(\tilde{U}_{d}^{\dagger}\right)_{j''j'} \left(\tilde{J}_{d}^{-1}\right)_{j'j},
\label{tildeyd}\\
&~& \tilde{y}_{ij}^{(e)} = \left(\tilde{J}_{l}^{-1}\right)_{ii'} 
\left(\tilde{U}_{l}\right)_{i'i''} \left\langle \left(Y_3\right)_{i''j''} \right\rangle
\left(\tilde{U}_{e}^{\dagger}\right)_{j''j'} \left(\tilde{J}_{e}^{-1}\right)_{j'j},
\label{tildeye}\\
&~& \tilde{y}_{ij}^{(\nu)} = \left(\tilde{J}_{l}^{-1}\right)_{ii'} 
\left(\tilde{U}_{l}\right)_{i'i''} \left\langle \left(Y_4\right)_{i''j''} \right\rangle
\left(\tilde{U}_{\nu}^{\dagger}\right)_{j''j'} \left(\tilde{J}_{\nu}^{-1}\right)_{j'j},
\label{tildeynu}
\end{eqnarray}
where $\displaystyle{\tilde{J}_{q}^{-1}}$, 
$\displaystyle{\tilde{J}_{u}^{-1}}$,
$\displaystyle{\tilde{J}_{d}^{-1}}$, 
$\displaystyle{\tilde{J}_{l}^{-1}}$,
$\displaystyle{\tilde{J}_{e}^{-1}}$,
and $\displaystyle{\tilde{J}_{\nu}^{-1}}$
are the inverse matrices of 
$\tilde{J}_q$, $\tilde{J}_u$, $\tilde{J}_d$, $\tilde{J}_l$, $\tilde{J}_e$, and $\tilde{J}_{\nu}$, respectively.

Third, we diagonalize $\left(\tilde{y}^{(u)}\tilde{y}^{(u)\dagger}\right)_{ij}$, 
$\left(\tilde{y}^{(d)}\tilde{y}^{(d)\dagger}\right)_{ij}$,
$\left(\tilde{y}^{(e)}\tilde{y}^{(e)\dagger}\right)_{ij}$, and
$\left(\tilde{y}^{(\nu)}\tilde{y}^{(\nu)\dagger}\right)_{ij}$ by unitary transformations as
\begin{eqnarray}
&~& \left(\tilde{V}_{\rm L}^{(u)}\right)_{ii'} 
\left(\tilde{y}^{(u)}\tilde{y}^{(u)\dagger}\right)_{i'j'}
\left(\tilde{V}_{\rm L}^{(u)\dagger}\right)_{j'j} = \left(\tilde{y}_{\rm diag}^{(u)2}\right)_{ij},~~
\label{tildeyuquark-diag}\\
&~& \left(\tilde{V}_{\rm L}^{(d)}\right)_{ii'} 
\left(\tilde{y}^{(d)}\tilde{y}^{(d)\dagger}\right)_{i'j'}
\left(\tilde{V}_{\rm L}^{(d)\dagger}\right)_{j'j} = \left(\tilde{y}_{\rm diag}^{(d)2}\right)_{ij},~~
\label{tildeydquark-diag}\\
&~& \left(\tilde{V}_{\rm L}^{(e)}\right)_{ii'} 
\left(\tilde{y}^{(e)}\tilde{y}^{(e)\dagger}\right)_{i'j'}
\left(\tilde{V}_{\rm L}^{(e)\dagger}\right)_{j'j} = \left(\tilde{y}_{\rm diag}^{(e)2}\right)_{ij},~~
\label{tildeyelepton-diag}\\
&~& \left(\tilde{V}_{\rm L}^{(\nu)}\right)_{ii'} 
\left(\tilde{y}^{(\nu)}\tilde{y}^{(\nu)\dagger}\right)_{i'j'}
\left(\tilde{V}_{\rm L}^{(\nu)\dagger}\right)_{j'j} = \left(\tilde{y}_{\rm diag}^{(\nu)2}\right)_{ij},~~
\label{tildeynulepton-diag}
\end{eqnarray}
where $\tilde{y}_{\rm diag}^{(u)2}$, $\tilde{y}_{\rm diag}^{(d)2}$, $\tilde{y}_{\rm diag}^{(e)2}$,
and $\tilde{y}_{\rm diag}^{(\nu)2}$ are 
$\tilde{y}_{\rm diag}^{(u)}$ squared, $\tilde{y}_{\rm diag}^{(d)}$ squared, 
$\tilde{y}_{\rm diag}^{(e)}$ squared, and
$\tilde{y}_{\rm diag}^{(\nu)}$ squared, respectively.

Last, we examine whether the following relations hold or not,
\begin{eqnarray}
\hspace{-0.5cm}&~& \left(\tilde{y}_{\rm diag}^{(u)}\right)_{ij} 
= \left(y_{\rm diag}^{(u)}\right)_{ij},~~
 \left(\tilde{y}_{\rm diag}^{(d)}\right)_{ij} = \left(y_{\rm diag}^{(d)}\right)_{ij},~~
 \left(\tilde{y}_{\rm diag}^{(e)}\right)_{ij} = \left(y_{\rm diag}^{(e)}\right)_{ij},~~
\label{tildey-check}\\
\hspace{-0.5cm}&~& \tilde{V}_{\rm L}^{(u)} \tilde{V}_{\rm L}^{(d)\dagger} = V_{\rm KM},~~
\tilde{V}_{\rm L}^{(e)} \tilde{V}_{\rm L}^{(\nu)\dagger} = V_{\rm MNS}.
\label{tildeV-check2}
\end{eqnarray}

\end{document}